\RequirePackage{rotating}
\documentclass[a4paper,fleqn,usenatbib]{mnras}

\usepackage[T1]{fontenc}
\usepackage{ae,aecompl}

\usepackage{graphicx}	
\usepackage{rotating}  
\usepackage{amsmath}	
\usepackage{amssymb}	
\usepackage{pifont}
\usepackage{caption} 

\title[Automated multi-band feature detection]{Generalised model-independent characterisation of strong gravitational lenses VIII. automated multi-band feature detection to constrain local lens properties}

\author[J.~Lin, J.~Wagner, and R.~E.~Griffiths]{
Joyce Lin$^1$\thanks{E-mail: joycelin2809@gmail.com}, Jenny Wagner$^{2}$\thanks{E-mail: thegravitygrinch@gmail.com} and Richard E.~Griffiths$^{1,3}$\thanks{E-mail: griff2@hawaii.edu}
\\
$^{1}$Department of Physics, Carnegie Mellon University, 5000 Forbes Avenue, Pittsburgh, PA 15213, USA, \\ 
$^{2}$Friedhofstr. 17, 66280 Sulzbach, \url{thegravitygrinch.blogspot.com}, \\ 
$^{3}$Department of Physics \& Astronomy, University of Hawaii at Hilo, 200 W. Kawili St, Hilo, HI 96720, USA
}

\date{Accepted XXX. Received YYY; in original form ZZZ}

\pubyear{2019}

\begin{document}
\label{firstpage}
\pagerange{\pageref{firstpage}--\pageref{lastpage}}
\maketitle

\begin{abstract}
As established in previous papers of this series, observables in highly distorted and magnified multiple images caused by the strong gravitational lensing effect can be used to constrain the distorting properties of the gravitational lens at the image positions. 
If the background source is extended and contains substructure, like star forming regions, which is resolved in multiple images, all substructure that can be matched across a minimum of three multiple images can be used to infer the local distorting properties of the lens.
In this work, we replace the manual feature selection by an automated feature extraction based on \texttt{SExtractor} for Python and show its superior performance.
Despite its aimed development to improve our lens reconstruction, it can be employed in any other approach, as well.
Valuable insights on the definition of an `image position' in the presence of noise are gained from our calibration tests.
Applying it to observations of a five-image configuration in galaxy cluster CL0024 and the triple-image configuration containing Hamilton's object, 
we determine local lens properties for multiple wavebands separately. 
Within current confidence bounds, all of them are consistent with each other, corroborating the wavelength-independence of strong lensing and offering a tool to detect deviations caused by micro-lensing and dust in further examples.
\end{abstract} 



\section{Introduction}
\label{sec:introduction}

Highly dense mass distributions like galaxy clusters or galaxies can act as strong gravitational lenses, deflecting the light of background objects into multiple images. 
Previously established approaches to reconstruct the deflecting mass density profiles, e.g.~\texttt{Lenstool} \citep{bib:Jullo_lenstool}, \texttt{Grale} \citep{bib:Liesenborgs_grale}, \texttt{Glee} \citep{bib:Grillo_glee}, or the light-traces-mass-approach of \citet{bib:Zitrin} to name a few galaxy-cluster reconstruction tools or, e.g.~the galaxy-scale reconstruction algorithms developed in \cite{bib:Keeton} and \cite{bib:Saha}, use global mass density profiles that are fit to the observables of the multiple images, mainly the positions of their centres of light.

The method developed in this paper series, mainly in paper~I \citep{bib:Wagner1}, paper~II \citep{bib:Wagner2}, and paper~III \citep{bib:Wagner3}, principally differs from this methodology.
It does not fit a \emph{global} mass density model to all multiple images together, but directly uses them to constrain \emph{local} properties of the gravitational lens at their positions and in their close proximity.
These local lens properties, ratios of scaled mass densities, reduced shear components and approximations to the critical curves, can be directly inferred from the equations of the gravitational lensing formalism without additionally imposing a mass density profile, may it be of parametric type like used in \texttt{Lenstool} or consisting of a basis function set as in \texttt{Grale}. 
Thus, the local lens properties yield all data-based information about the lens that the model-based approaches have in common. 
Using these local lens properties, we can also back-project the multiple images into the source plane to obtain a reconstruction of the source morphology.
The latter can be determined up to an overall scaling factor and is free of any bias that may be caused during the back-projection when using a mass density profile model.
As detailed in paper~I, II, and III of this series, the extended surface brightness profiles of multiple images deliver the most constraining observables to apply this method.

Paper~IV \citep{bib:Wagner4} and VI \citep{bib:Wagner6} of the series deal with the mathematical derivation and the physical interpretation of the degeneracies of the strong gravitational lensing formalism.
They show which transformations can be applied to the equations of the lensing formalism leaving all observables in the multiple images invariant. 
In paper VII \cite{bib:Wagner7} the degeneracies between local lens properties, in particular higher-order terms like flexion, and intrinsic source characteristics, such as ellipticity, are investigated further.
Since paper~IV and VI investigated line-of-sight contributions to the degeneracies as well, we used a supernovae ensemble to set up observation-based cosmic distances that are least model-dependent in paper~V \citep{bib:Wagner5}. 
These are usually needed to determine the distances between us as observers, the lens, and the source.
A summary of the theory and some applications of the method highlighting its usefulness to track dark matter can be found in \cite{bib:Wagner_summary}.

This Paper~VIII aims to set up a robust, automated image processing pipeline in order to identify the peculiarities in surface brightness profiles that are used as observables to constrain local lens properties as detailed in paper~I and II. 
Due to the increasing amount and quality of multiple-image observations, several use-cases for our approach have already been discovered. 
On galaxy-cluster scale, we applied it to the five-image configuration of a blue spiral background galaxy in CL0024 in \cite{bib:Wagner_cluster} and recently to the triple-image configuration of a star-forming galaxy in a newly detected galaxy cluster as detailed in \cite{bib:Griffiths}.  
For these two cases, we identified matching brightness features in multiple images manually by eye.
Only in \cite{bib:Wagner_quasar}, we could use pre-determined brightness features which were provided by a standard analysis tool of radio interferometer data.  
With many more possible application cases being discovered, we now address the issue how to extract and employ quantitatively robust features that can be detected automatically using image processing algorithms for given spatial distributions of light, like optical filter bands or radio band data after the conversion into the spatial domain.

Emphasis is put on an automated, quantitatively measurable feature extraction and not on an automated feature-matching process across multiple images because the former is independent of any strong gravitational lens modelling and therefore allows to infer local lens properties devoid of assumptions on the global deflecting mass density. 
As detailed below, this is also the reason why the feature extraction can be based on a multi-purpose, standard astrophysical image processing tool.
Subsequently matching these features automatically would require the use of additional assumptions or models as in lens reconstruction algorithms mentioned above. 
Thus, local lens-model-independent properties cannot be obtained in this way, which is the reason why we only aim for an automated feature extraction.

For the sake of self-consistency, Section~\ref{sec:theoretical_background} briefly summarises the approach to determine local lens properties out of substructure in extended surface brightness profiles of multiple images in the theoretical framework of single-plane strong gravitational lensing. 
In Section~\ref{sec:robust_feature_extraction}, we detail our data processing pipeline from the data used to the extraction of our features of interest which are to be inserted into the equations introduced in Section~\ref{sec:theoretical_background}.  
As many well-established collaborations have relied on \texttt{SExtractor} \citep{bib:Bertin} as the standard software to generate catalogues of multiple-image observables, like \cite{bib:Postman} or \cite{bib:Lotz}, we also base our data processing on it. 
Yet, instead of using \texttt{SExtractor} directly, we employ the routines provided by the `Python library for Source Extraction and Photometry', \texttt{sep}, \citep{bib:Barbary} because recent developments in cosmology and astronomy increasingly rely on Python as the standard programming language. 

After the introduction of the robust feature extraction for our approach, Section~\ref{sec:applications} shows the resulting local lens properties based on these improved features for the two previously mentioned multiple-image configurations and compares them to the results obtained with the features identified by eye in \citet{bib:Wagner_cluster} and \citet{bib:Griffiths}.  
Beyond the comparison of performance between manual and automated feature extraction, we also investigate the dependence of the local lens properties on the filter band that is used to extract the features. 
As gravitational lensing effects are wavelength-independent, analysing the same configuration over multiple filter bands allows us to investigate biases, for instance due to dust extinction.
Subsequently, Section~\ref{sec:discussion} summarises the results and discusses the advantages of the new robust feature extraction for our lens reconstruction approach and others.
Section~\ref{sec:conclusion} concludes the paper discussing further usages to robustly detect deviations from wavelength-independent strong gravitational lensing in further multiple-image configurations.

\section{Theoretical background}
\label{sec:theoretical_background}

A general overview of gravitational lensing in arbitrary spacetimes can be found in \cite{bib:Fleury2} or any standard text book on gravitational lenses, for instance, \cite{bib:SEF}.
The approach discussed here is based on the standard single-lens-plane strong gravitational lensing formalism in a concordance $\Lambda$-Cold-Dark-Matter cosmological background. 

For consistency across all papers of this series, we denote two-dimensional positions in the lens plane by $\boldsymbol{x}$.
The angular diameter distance to the lens plane is $D_\mathrm{d}$ corresponding to its redshift $z_\mathrm{d}$. 
Analogously, we denote two-dimensional positions in the source plane by $\boldsymbol{y}$, the angular diameter distance is $D_\mathrm{s}$ corresponding to $z_\mathrm{s}$. 
The distance between lens and source plane is denoted by $D_\mathrm{ds}$.
Multiple images are indexed by $i,j = 1, 2, 3, ...$ and positions of brightness features by $\mu = 1, 2, 3, ...\,$. 

\subsection{Summary of the formalism to extract local lens properties}
\label{sec:summary_formalism}

\begin{figure}
\centering
\includegraphics[width=0.45\textwidth]{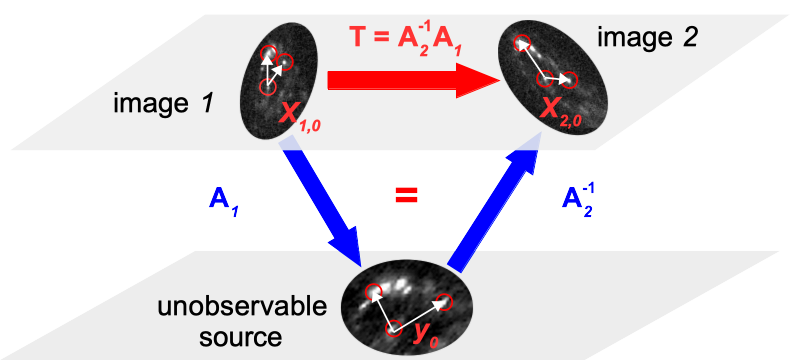}
\caption{Visualisation of the \texttt{ptmatch} principle: corresponding brightness features (red circles) are identified in multiple images, so that the system of equations of Eq.~\eqref{eq:matching} determines the local lens properties in Eqs.~\eqref{eq:f}, \eqref{eq:g}, and \eqref{eq:j}. As the graphics shows, a linear transformation $\rm{T}$ between two images close to corresponding multiple-image positions $\boldsymbol{x}_{1,0}$, $\boldsymbol{x}_{2,0}$ can be determined from the brightness feature vectors (white arrows). This $\rm{T}$ corresponds to the product of the distortion matrices of these multiple images, enabling \texttt{ptmatch} to solve for the local lens properties.}
\label{fig:ptmatch}
\end{figure}

For self-consistency, we briefly summarise the approach of \cite{bib:Wagner2} and \cite{bib:Wagner_cluster} to match features in multiple images onto each other in order to extract local lens properties. 
The linearised lens equation around a point $\boldsymbol{x}_{i,0}$ in multiple image $i = 1, 2, 3, ...$ maps vectors around $\boldsymbol{x}_{i,0}$ into vectors in the source plane around the source point $\boldsymbol{y}_{0}$, which is the common source position of the multiple image points $\boldsymbol{x}_{i,0}$. 
Thus, using the distortion matrices ${\rm{A}}(\boldsymbol{x}_{i,0})$, we can state
\begin{equation}
\boldsymbol{y}_{\mu} - \boldsymbol{y}_{0} = {\rm{A}}(\boldsymbol{x}_{i,0}) \left(\boldsymbol{x}_{i,\mu} - \boldsymbol{x}_{i,0} \right) = {\rm{A}}(\boldsymbol{x}_{j,0}) \left(\boldsymbol{x}_{j,\mu} - \boldsymbol{x}_{j,0} \right) \;,
\label{eq:matching}
\end{equation}
meaning that the lens mapping back-projects vectors between corresponding brightness features $\mu=1,2,...$ in multiple images, here image~$i$ and image~$j$, onto the same vector in the source plane (see also Fig.~\ref{fig:ptmatch}). 

The distortion matrix usually contains the convergence $\kappa$ and the shear components $\gamma_1$ and $\gamma_2$ at the position $\boldsymbol{x}_{i,0}$.
The former is a measure of the local scaled mass density which scales the source properties by an overall factor. 
The latter two represent the local leading-order distorting strength of the lens.
Yet, as we are interested in properties that can be directly linked to observables, we write ${\rm{A}}_i \equiv {\rm{A}}(\boldsymbol{x}_{i,0})$ in terms of the reduced shear components $g_1$ and $g_2$ as
\begin{equation}
{\rm{A}}_i = \left( 1 - \kappa \right) \left( \begin{matrix} 1 - g_1 & -g_2 \\ -g_2 & 1 + g_1 \end{matrix} \right) \;.
\label{eq:A}
\end{equation} 

Since the vectors in the source plane cannot be observed, we first identify corresponding features in the multiple images to set up the vectors $\left(\boldsymbol{x}_{i,\mu} - \boldsymbol{x}_{i,0} \right)$ for $i=1, 2, 3, ...$.
Subsequently, we use the last part of Eq.~\eqref{eq:matching} to set up a system of equations linking the vectors in different multiple images and their unknown distortion matrices, which is then solved for the ratios of convergences between multiple images
\begin{equation}
f_i = \dfrac{1-\kappa_1}{1-\kappa_i} \;, \quad i = 2, 3, ...
\label{eq:f}
\end{equation}
and the reduced shear components at all multiple image positions
\begin{equation}
g_{i,1} = \dfrac{\gamma_{i,1}}{1-\kappa_i} \;, \qquad g_{i,2} = \dfrac{\gamma_{i,2}}{1-\kappa_i} \;, \quad i=1, 2, 3, ...
\label{eq:g}
\end{equation}
with the magnitude of the shear of image~$i$ given by
\begin{equation}
	|\boldsymbol{g}|_i = \sqrt{g_{i,1}^2 + g_{i,2}^2}\;, \quad i=1,2,3, ...\;,
	\label{eq:g_magnitude}
\end{equation}
if we have at least three multiple images each containing two non-parallel vectors that can be matched. 
Fig.~\ref{fig:ptmatch} visualises the matching, \cite{bib:Wagner2} details the minimum requirements and degeneracies when solving the system of equations and \cite{bib:Wagner_cluster}  details the implementation. 
From Eqs.~\eqref{eq:f} and \eqref{eq:g} using Eq.~\eqref{eq:A}, we also determine the ratios of magnification ratios
\begin{equation}
\mathcal{J}_i = \left(\det \left({\rm{A}}_i\right) \right)^{-1} \det \left({\rm{A}}_1\right) \;, \quad i=2,3, ...\;, 
\label{eq:j}
\end{equation}
which can be compared to observed flux ratios between the multiple images for a consistency check. 
Differences between Eq.~\eqref{eq:j} determined by the brightness feature locations and the observed flux ratios can hint at additional micro-lensing of individual images or dust extinction due to intervening gas clouds along the line of sight. 

Only ratios of the characteristic local lens properties $\kappa$, $\gamma_1$, $\gamma_2$ can be determined because all observables are angular distances on the celestial sphere (see \cite{bib:Wagner4} and \cite{bib:Wagner6} that these quantities are not subject to any of the common strong-lensing degeneracies and that they are indeed the maximum information retrievable from observables without imposing additional model assumptions).
To be converted to physical distances, they require an overall size scale, like a physical distance, to be fixed. 
In standard approaches to gravitational lens reconstructions, the angular diameter distances based on a chosen cosmological model or on observations, like set up for our approach in \cite{bib:Wagner5}, are used.  

In principle, Eq.~\eqref{eq:f} can contain any pair of convergences, but, for the sake of efficiency, the implementation we use, called \texttt{ptmatch}\footnote{Available at \url{https://github.com/ntessore/imagemap}.}, fixes a so-called `reference image' with respect to which all ratios of convergences are determined. 
As in all related works, we set the reference image to be called image~$1$. 
\texttt{ptmatch} also takes into account uncertainties in the brightness feature positions and yields confidence bounds on the local lens properties of Eqs.~\eqref{eq:f} and \eqref{eq:g} as sampled from the full likelihood distribution of the $f$ and $\boldsymbol{g}$ values for the given observed positions. 
Details about the confidence bounds can also be found in \cite{bib:Wagner2} and \cite{bib:Wagner_cluster}.

\subsection{Robust features}
\label{sec:robust_features}

Based on the strong lensing formalism, the most reasonable robust features are maxima in the brightness distribution of the multiple images because maximum intensities in the surface brightness profile of the source object are mapped into maximum intensities in the surface brightness profiles of the multiple images. 
This statement relies on the conservation of photons and can be proven as follows:
Assume the intensity at a point $\boldsymbol{y}_0$ in the source object is $I(\boldsymbol{y}_0)$.
If this is an intensity maximum, then
\begin{equation}
\nabla_{\boldsymbol{y}} I(\boldsymbol{y}_0)=0
\label{eq:maximum}
\end{equation}
holds at $\boldsymbol{y}_0$, meaning, we first take the gradient $\nabla_{\boldsymbol{y}}$ with respect to $\boldsymbol{y}$ in the source plane and then evaluate it at $\boldsymbol{y}_0$.
Now, let $\boldsymbol{x}_0$ be a corresponding image point to $\boldsymbol{y}_0$ in the lens plane with intensity $I(\boldsymbol{x}_0)$, like, for instance, $\boldsymbol{x}_{1,0}$ to $\boldsymbol{y}_0$ in Fig.~\ref{fig:ptmatch}. 
If the surface brightness is conserved, then $I(\boldsymbol{x}_0) = I(\boldsymbol{y}_0)$.
Deriving both sides with respect to $\boldsymbol{x}$ implies
\begin{equation}
\nabla_{\boldsymbol{x}} I(\boldsymbol{x}_0) = \nabla_{\boldsymbol{x}} I(\boldsymbol{y}_0) = A(\boldsymbol{x}_0) \nabla_{\boldsymbol{y}} I(\boldsymbol{y}_0) = 0 \;,
\label{eq:maximum_conservation}
\end{equation}
where we transformed the gradient from the lens to the source plane first and then used Eq.~\eqref{eq:maximum} in the last step. 
Although the photon conservation is an idealised assumption that may not strictly hold for observations, cross checks can be made when jointly analysing data from several filter bands and using Eq.~\eqref{eq:j} (see detailed discussions in \cite{bib:Wagner_cluster, bib:Wagner_quasar, bib:Griffiths} for three examples, differently affected by observational biases like stray light and extinction). 

Since Eq.~\eqref{eq:maximum_conservation} also implies that intensity minima in the source plane are mapped to intensity minima in the lens plane, one could also think of using them as `darkness features'.
Some features used in \cite{bib:Griffiths} actually relied on this argument and yielded a robust and reasonable \texttt{ptmatch} result. 
However, the darkness features may remain rather an exception than becoming standard because the negligible contaminations due to stray light of neighbouring foreground objects and the very low and stable background noise level as observed for the triple-image configuration in \cite{bib:Griffiths} has only been rarely found so far.

\section{Robust feature extraction by image processing}
\label{sec:robust_feature_extraction}

\begin{figure*}
  \centering
  \includegraphics[width=\linewidth]{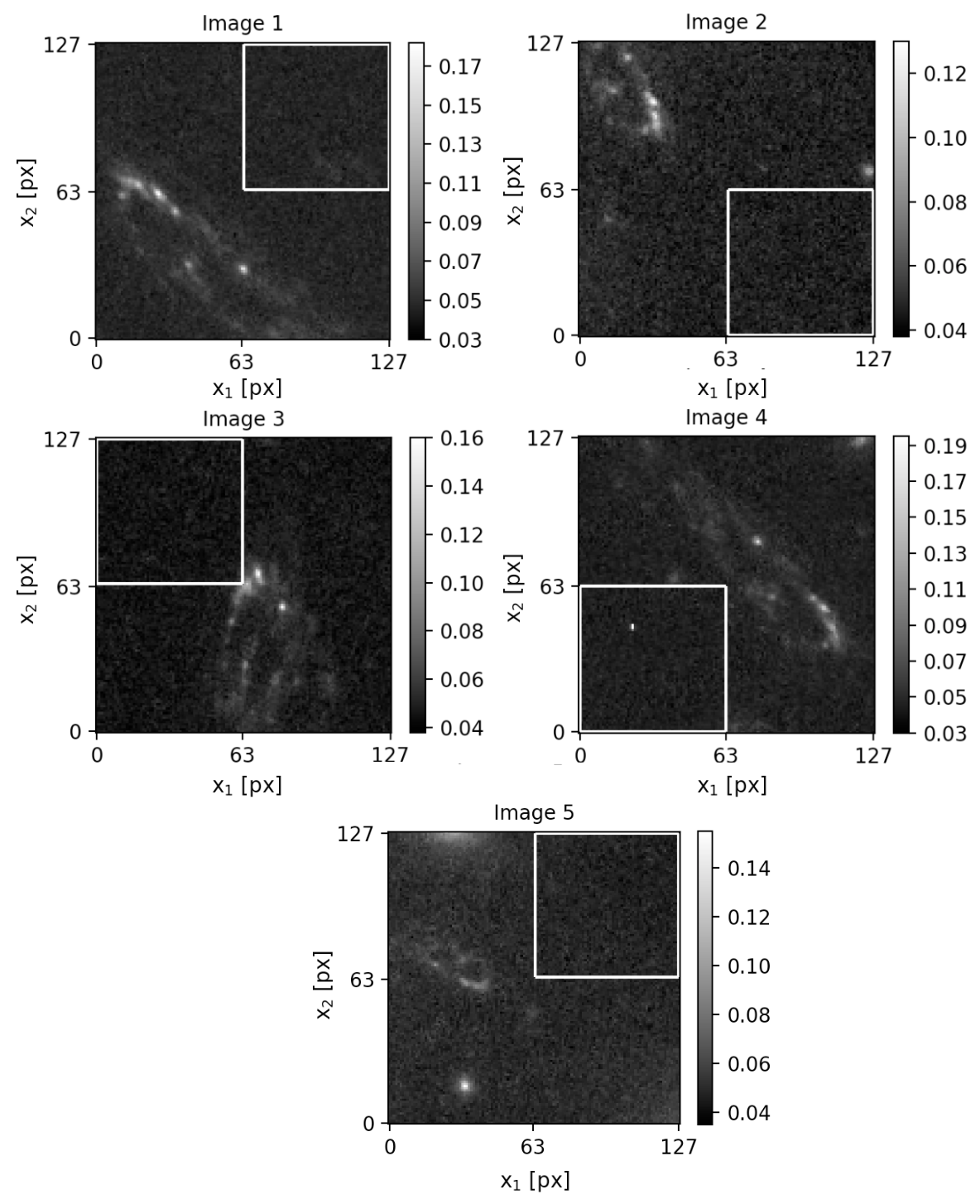}
  \caption{Original data used in Section~\ref{sec:robust_feature_extraction} to set up the robust feature extraction: Image~1--5 in the cluster-scale lens CL0024 in the HST ACS/WFC F475W filter band in $128~\times~128$ boxes including $64~\times~64$ patches (white squares) for background estimation.
  }
  \label{fig:Image_bkg_CL0024}
\end{figure*}

As motivated in Section~\ref{sec:introduction}, our image processing pipeline to extract the intensity maxima as robust features, detailed in Section~\ref{sec:robust_features}, relies on the `Python library for Source Extraction and Photometry', \texttt{sep}, and consists of the usual steps of background subtraction (see Section~\ref{sec:background_subtraction}) and subsequent feature detection and extraction (see Section~\ref{sec:feature_extraction}).
We systematically analyse the optimum parameter ranges for both subroutines to maximise the number and stability of detected intensity maxima for the star-forming regions of the multiple-image configuration in CL0024 discussed in \citet{bib:Wagner_cluster} in the 475W filter band. 
We will show in this and the following section that training and calibrating on a single, very clear example case in one filter band, generalises well to further observations of other wavelengths because all steps are based on sound physical and image-processing principles.
Limitations, as also noted by other automated lens reconstruction pipelines like \cite{bib:Shajib}, are summarised in Section~\ref{sec:conclusion}. 

To accelerate the image processing, we choose a $128~\times~128$ pixel box to contain the individual gravitationally lensed images, including an area of $64~\times~64$ pixels devoid of any image pixels and further unrelated high-intensity signals for the local background estimation. 
Fig.~\ref{fig:Image_bkg_CL0024} shows the boxes (with areas for background estimation marked by white squares) for all five images of our example used in this section.  
No pixels were excluded from the analysis pipeline, meaning that we did not make use of any masking functions due to the negligible number of corrupt pixels. 
As detailed below, these choices are applicable to many other multiple images as observed with current telescopes, in particular the Hubble Space Telescope (HST).

\subsection{Background subtraction}
\label{sec:background_subtraction}

The background subtraction routine of \texttt{sep} is based on the same mixture of $\kappa$ $\sigma$ clipping and mode estimation as \texttt{SExtractor}, documented in \citet{bib:Bertin}. 
As such, it could estimate a varying background over an extended region observed on the sky even in the presence of foreground signals and stray light. 
However, run-time is increased when applying this routine to an extended region. 
Furthermore, the robustness of the resulting background estimate underneath the foreground pixel has to be checked, see, for instance, \citet{bib:Melchior3} for a more detailed discussion. 
We therefore choose to estimate a constant mean background value, $\mu_{\rm{bg}}$ and its root-mean-square value, $\sigma_{\rm{I}}$, from a sufficiently large, separate background patch close to the foreground pixels of the gravitationally lensed images (see white squares in Fig.~\ref{fig:Image_bkg_CL0024}). 

Using a patch size of $64~\times~64$ (the \texttt{sep} default value) approximately covers the same area spanned by the pixel of a gravitationally lensed image at HST resolution, so that the resulting $\sigma_{\rm{I}}$ obtained from this size of background patch yields an adequate measure of the noise level per pixel to robustly detect intensity maxima as detailed in Section~\ref{sec:feature_extraction}. 
Using smaller or larger patch sizes can lead to a reduced amount of detected features or a less efficient detection, depending on whether the respective $\sigma_{\rm{I}}$ is larger or smaller than the optimum value\footnote{
Assuming the $\kappa$ $\sigma$ clipping and mode estimation to work perfectly in an idealised setting, increasing the number of analysed pixel would converge to the statistical noise level of the data acquisition.
}. 
Accounting for potentially occurring biases in the background estimation, we define the sufficient ($\mu_{\rm{bg}}$, $\sigma_{\rm{I}}$) pair when the maximum intensity positions of all star-forming regions at significance levels larger or equal than $3\sigma_{\rm{I}}$ above $\mu_{\rm{bg}}$ are correctly identified. 
Choosing the background patch size too small or too large particularly affects the quality of feature detection in multiple images in the central part of galaxy clusters, as image~5 in Fig.~\ref{fig:Image_bkg_CL0024} demonstrates.
Too small a background patch size will set $\mu_{\rm{bg}}$ as too high, while too large a background patch will increase $\sigma_{\rm{I}}$ when extending the patch out to less luminous areas in the galaxy cluster. 
As a consequence, the features with low signal to noise in these images will remain undetected. 

Next, we determine a suitable value for the width and height of the median filtering which is typically used to suppress bright foreground signals like stars. 
As expected by construction of background patches, different values of the filter width and height do not change the detected number of features (see Section~\ref{sec:feature_extraction} for details on the detection and extraction step), so that we keep the default value of three pixel. 

Thus, the parameter settings for the \texttt{sep} background estimate routine are set to the default values having corroborated their usefulness for our case. 
After applying this routine to all background patches shown in Fig.~\ref{fig:Image_bkg_CL0024}, we retrieve the \emph{globalback} and \emph{globalrms} return values.
They correspond to $\mu_{\rm{bg}}$ and $\sigma_{\rm{I}}$, respectively, so that we then can subtract the mean background estimate from the entire $128~\times~128$ box to continue with the feature identification in the multiple images.

\subsection{Feature extraction}
\label{sec:feature_extraction}

To identify and extract the maximum intensity values of the star-forming regions, we continue with the \texttt{sep} routine to extract these features. 
The latter requires a background-subtracted array of data as input. 
We insert the resulting $128~\times~128$ pixel box obtained after the steps detailed in Section~\ref{sec:background_subtraction}. 

\begin{figure}
  \centering
  \includegraphics[width=0.45\textwidth]{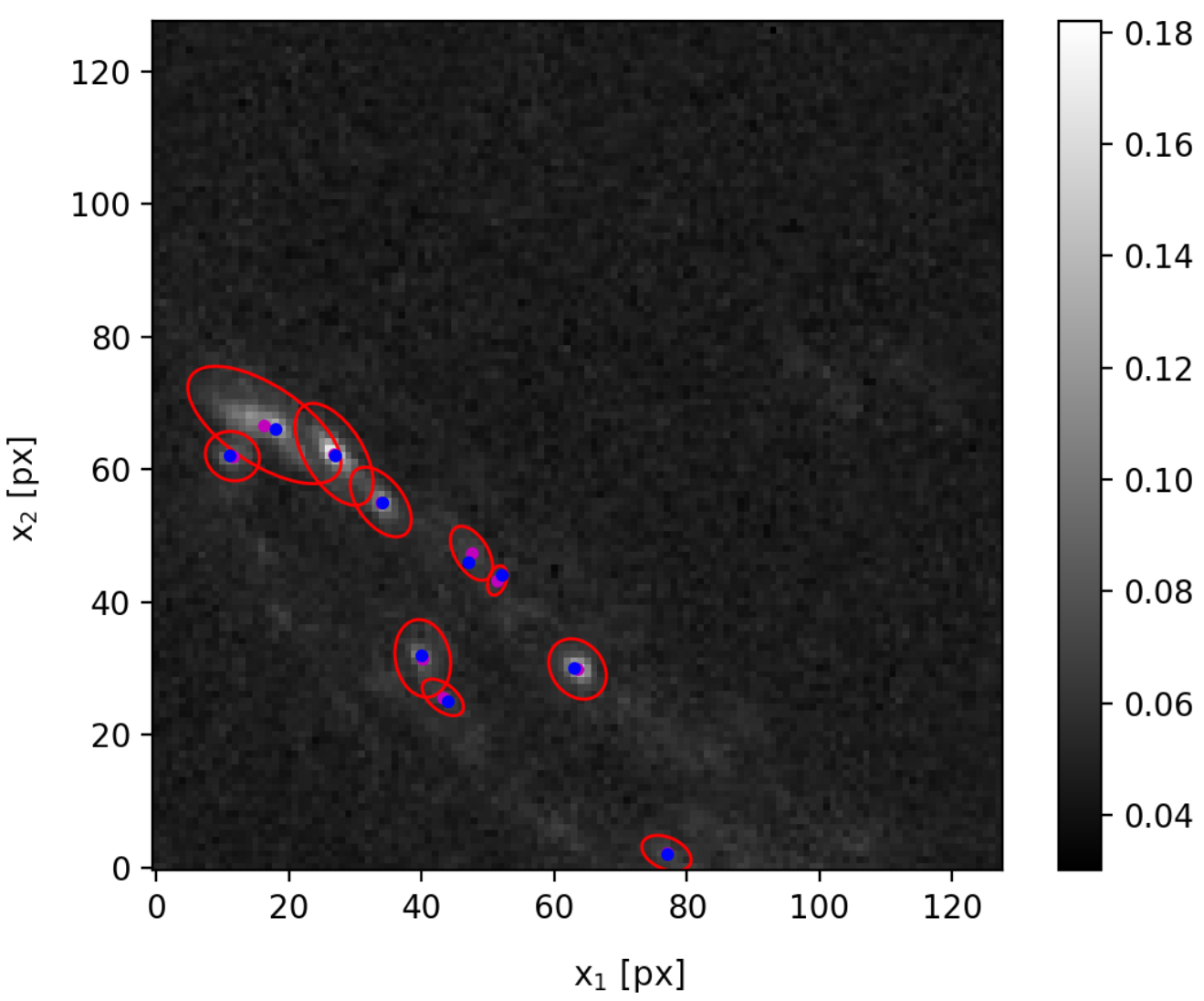}
  \caption{Object detection for the pixel box around image~1 of Fig.~\ref{fig:Image_bkg_CL0024} for $n_{\rm{I}}=5$. Detected objects are marked by their centre of light (magenta dot), the pixel of maximum intensity (blue dot), and the extensions of their intensity quadrupole (red ellipses).}
  \label{fig:Image1_sig5}
\end{figure}

To separate foreground from background objects, \texttt{sep} employs a simple thresholding at each pixel
\begin{equation}
  t = \mu_{\rm{bg}} + n_{\rm{I}} \cdot \sigma_{\rm{I}} \;,
\label{eq:threshold}
\end{equation}
in which the signal strength $n_{\rm{I}}$ is user-defined and ($\mu_{\rm{bg}}$, $\sigma_{\rm{I}}$) are determined by the routine of Section~\ref{sec:background_subtraction}. 
Contiguous areas of pixels above $t$ are subsequently grouped into foreground objects. 
(The significance of detection in terms of $n_{\rm I} \sigma_{\rm I}$ above background thus marks the pixel-wise signal strength and should not be confused with the integrated signal strength for the whole object based on all of its pixels.)
The resulting collection of detected objects and their properties can be visualised as shown for the example of image~1 in Fig.~\ref{fig:Image1_sig5} for $n_{\rm{I}}=5$. 

\begin{figure}
  \centering
  \includegraphics[width=0.45\textwidth]{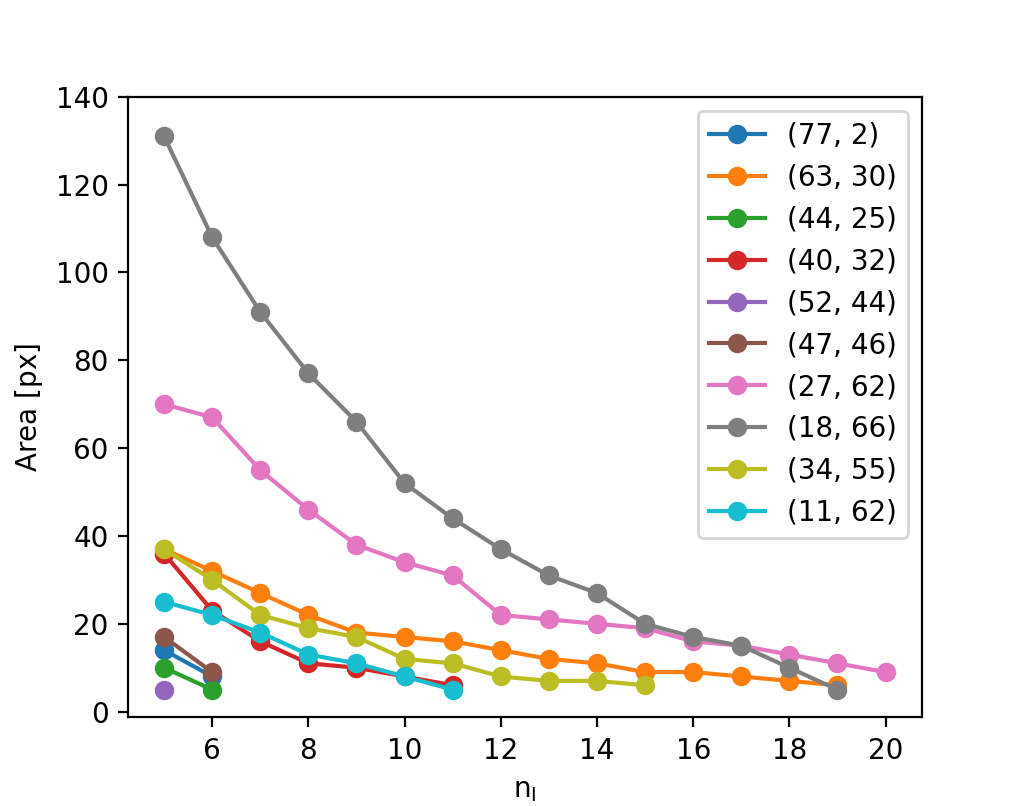}
  \caption{Dependence of the area of detected objects in image~1 (see Fig.~\ref{fig:Image1_sig5}) on the detection threshold, increasing $n_{\rm I}$ from 3 in steps of 1 until only a single object remains. The area is measured by the number of pixel assigned to the object and objects are denoted by the location of their peak position $(x_1,x_2)$.
  }
  \label{fig:area_vs_sigma}
\end{figure}

Increasing $n_{\rm{I}}$ from 3 in steps of 1 to the value at which no objects are detected anymore, we track the size of the detected objects
in the box surrounding image~1 of Fig.~\ref{fig:Image_bkg_CL0024} as plotted in Fig.~\ref{fig:area_vs_sigma}. 
The plot shows that the star-forming regions of interest can be extracted in this way because many of them are identified as objects up to a much larger threshold than set by the minimum requirement of $n_{\rm I}=3$. 
To avoid spurious detections, the minimum amount of required pixels to be counted as an object, \emph{minarea}, is kept at the default of five pixel. 
While the star-forming regions of interest cover a much larger area at $n_{\rm I}=3$--5, the next step of our robust feature extraction goes to the limit of higher thresholds and smaller areas. 
Therefore, we do not choose to set \emph{minarea} to a higher value.

As star-forming regions may lie closely together or be even merged into elongated wingy objects, we need to switch the \emph{clean} parameter in the \texttt{sep} extraction routine off to assure that all star-forming regions can be detected and are not pruned as spurious detection in the wings of a brighter one. 

\begin{figure}
  \centering
  \includegraphics[width=0.49\textwidth]{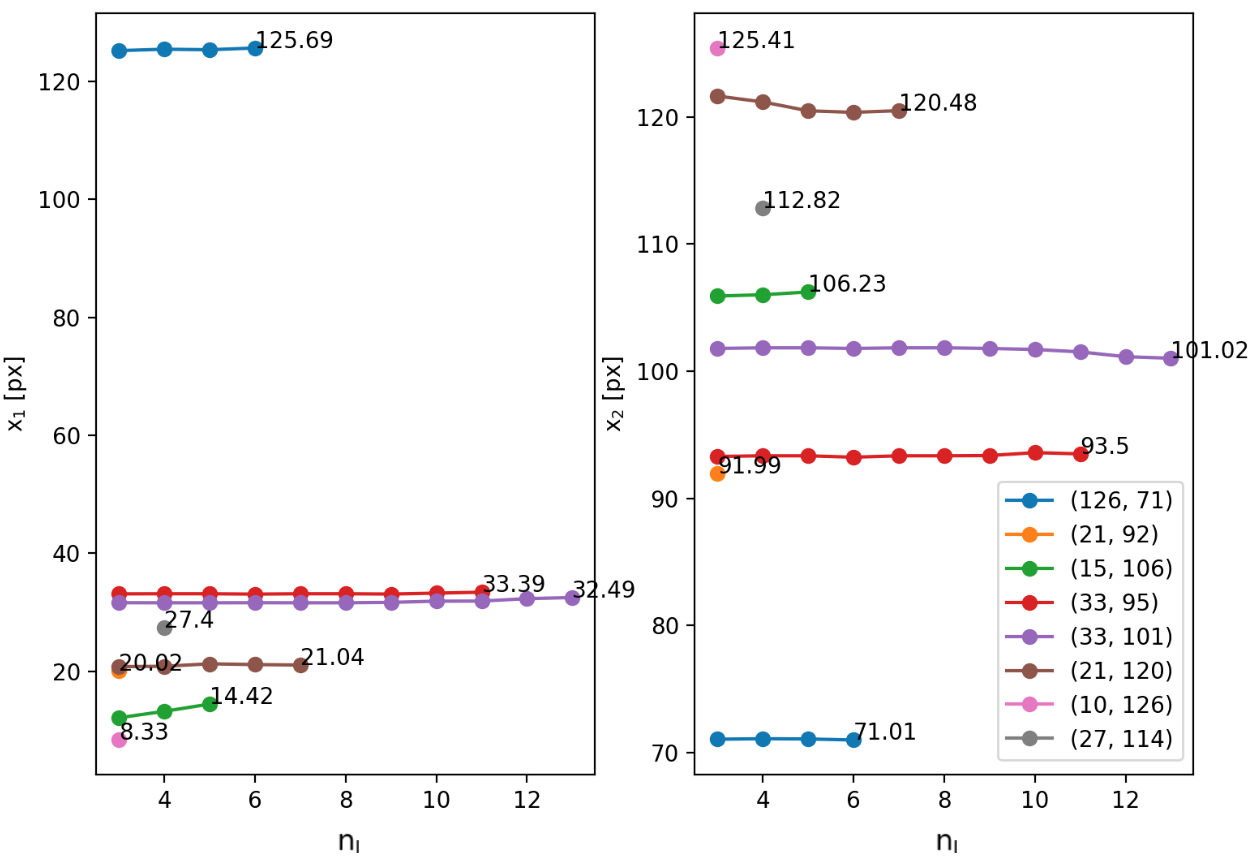}
  \vspace{-3ex}
  \caption{Persistence diagram for objects detected in image~2 of Fig.~\ref{fig:Image_bkg_CL0024}, centre of light positions in $x_1$ (left) and $x_2$ (right) over increasing threshold in intensity. The \emph{deblend\_cont} value is set to 0.005, so that all features persist over increasing $n_{\rm I}$. Numbers at the last entry denote the centre of light position at this threshold. Each object is denoted by its peak position in the legend.
  }
  \label{fig:Image_2_loc}
\end{figure}

To further maximise the number of features that can be matched across all multiple images, we tune the deblending parameter to split a multi-modal intensity distribution into separate objects. 
Tracking the creation and annihilation of objects is easiest in a `persistence diagram', as shown in Fig.~\ref{fig:Image_2_loc} for the example of image~2 of Fig.~\ref{fig:Image_bkg_CL0024}. 
The left-hand side tracks the $x_1$-coordinate of all objects for increasing threshold, the right-hand side the $x_2$-coordinate. 
Objects below five pixel area simply cease to exist, while new ones can be created at higher thresholds when large contiguous areas split into separate parts. 

\begin{figure}
  \centering
  \includegraphics[width=0.49\textwidth]{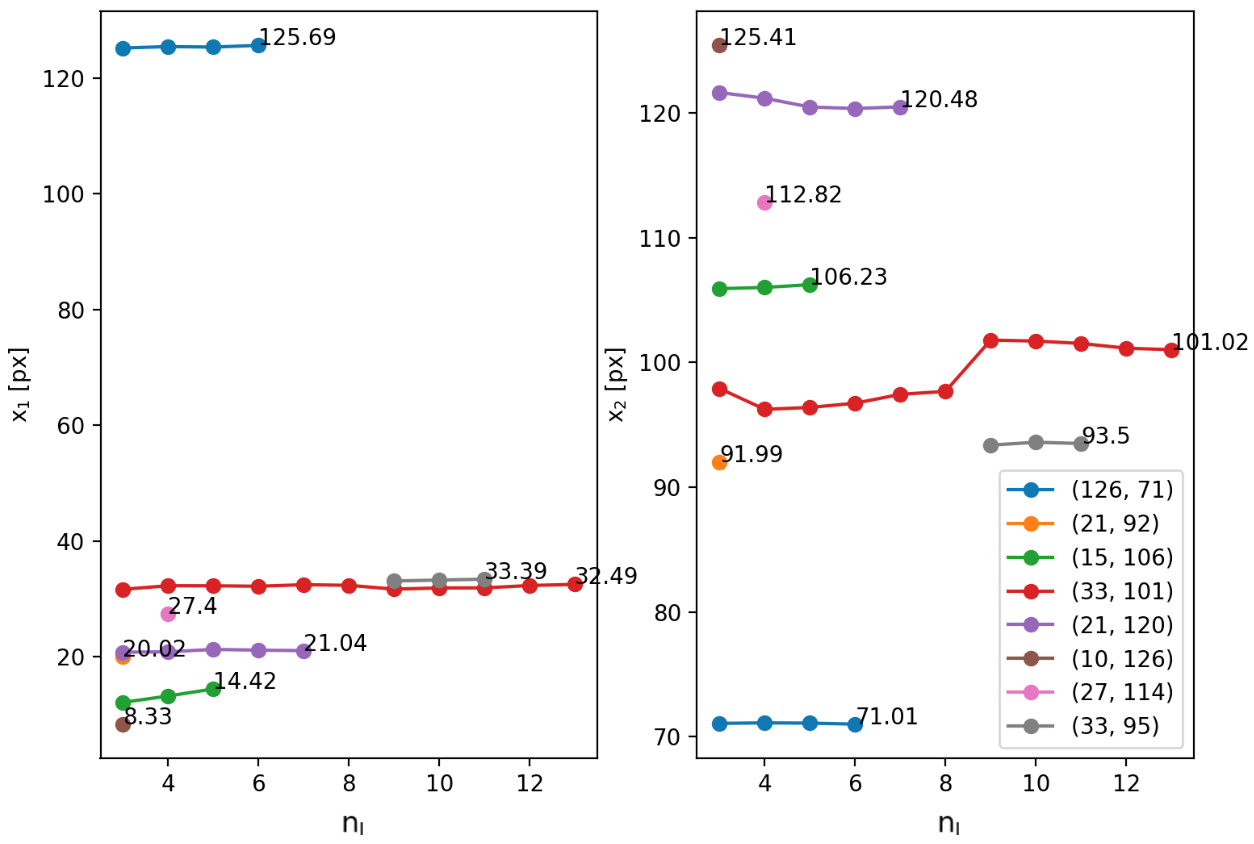}
  \vspace{-3ex}
  \caption{Same as Fig.~\ref{fig:Image_2_loc} but for \emph{deblend\_cont} set to 0.1, which clearly shows merging objects and thus no clear separation of features of interest anymore. A further indicator of instability is the difference between the last centre of light coordinates and the peak position. 
  }
  \label{fig:Image_2_deb_0.1}
\end{figure}

Fig.~\ref{fig:Image_2_loc} is plotted setting the minimum contrast ratio used for object deblending, \emph{deblend\_cont}, to its default value 0.005.
Hence, the minimum flux ratio to split a fainter object with lower flux from a brighter one with larger flux is 0.005.
The positions of all objects detected in this way remain stable over all thresholds and only one object is newly created above $n_{\rm I}=3$. 
In contrast, we increase \emph{deblend\_cont} to 0.1 in Fig.~\ref{fig:Image_2_deb_0.1} and observe that detected object positions become less stable.
Most severely the objects with coordinates $(33,95)$ and $(33,101)$ are affected, which, due to their mutual proximity, are not reliably separated anymore. 
Testing \emph{deblend\_cont} values from 0.000 to 1.000, we find most stable object separation for the default value, without introducing spurious artefacts at values below 0.005. 
As expected, increasing \emph{deblend\_cont} above 0.005, causes objects to mix and the number of detected objects to decrease because faint features disappear and features that lie close together merge.

Thus, the only change in the default setting of the \texttt{sep} extraction routine is to set the \emph{clean} parameter off, all other pre-defined defaults are suitable for our application. 

\begin{figure*}
  \centering
  \includegraphics[width=0.4\textwidth]{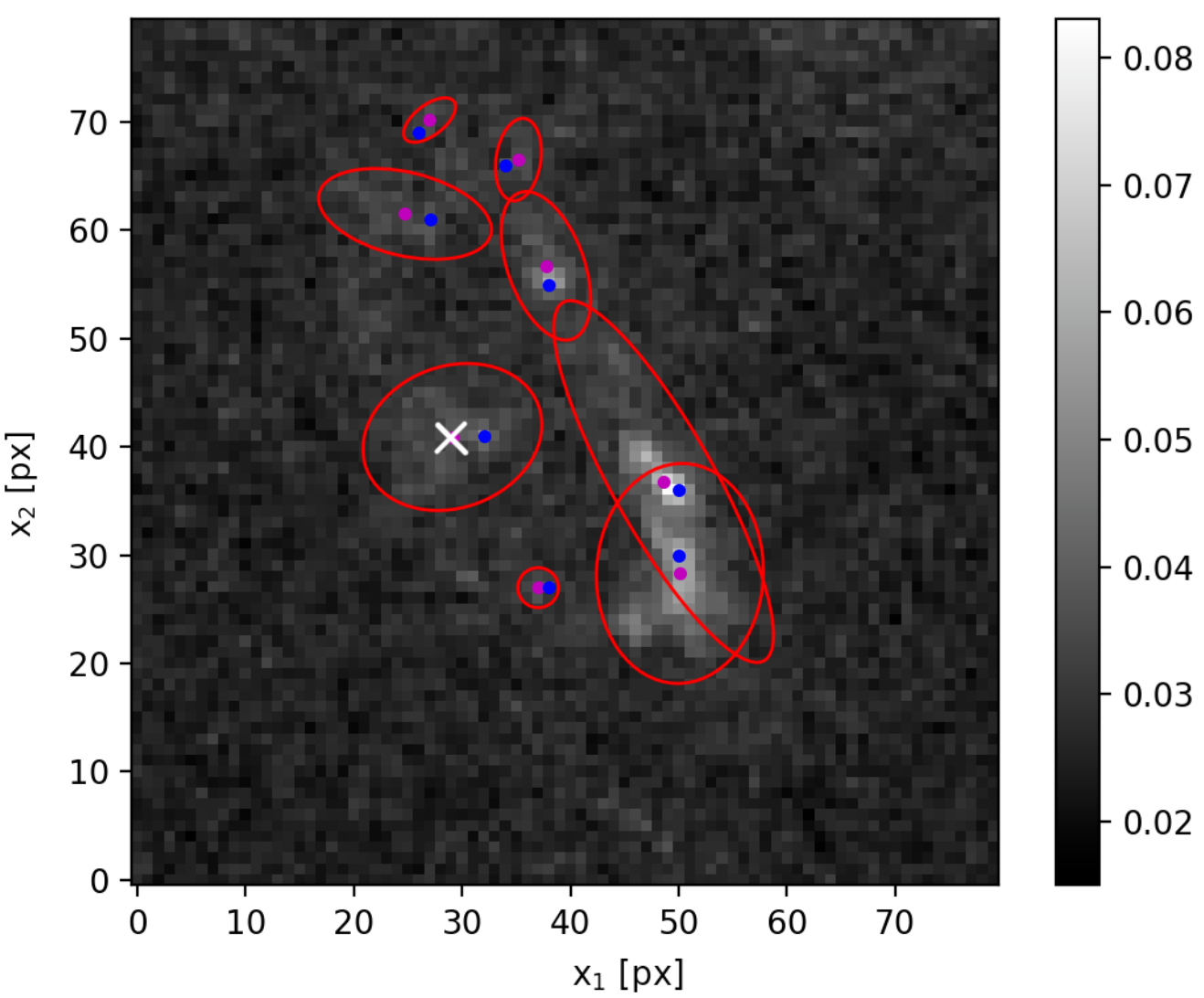}
  \hspace{0.1\textwidth}
  \includegraphics[width=0.4\textwidth]{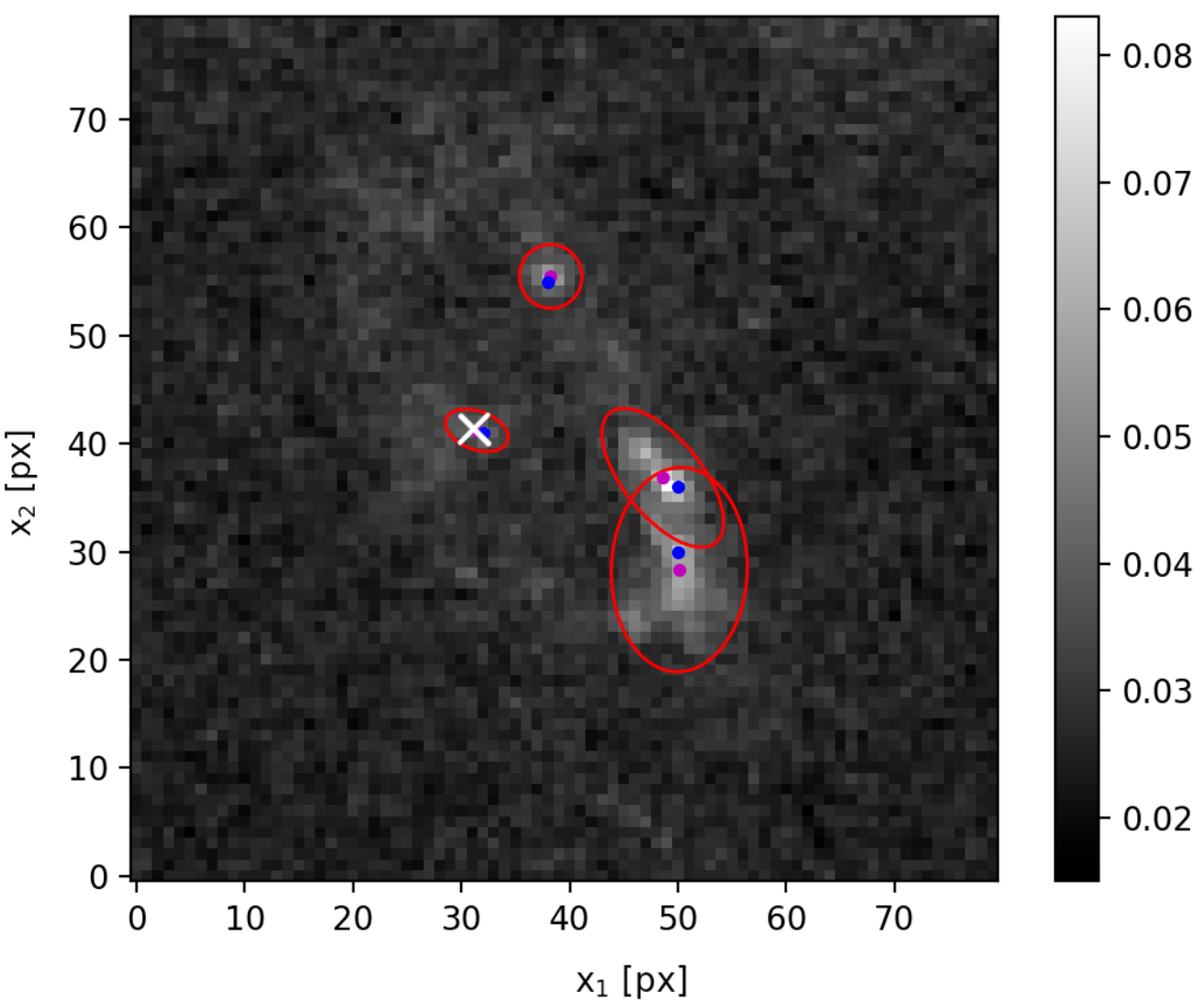}
  \caption{Robust feature selection from objects in image~2 of Fig.~\ref{fig:Image_bkg_CL0024}. Object detection and properties as in Fig.~\ref{fig:Image1_sig5}, coordinates deviate from Fig.~\ref{fig:Image_bkg_CL0024} for better visualisation. A comparison of detected objects at $n_{\rm I}=3$ (left) with those detected at $n_{\rm I}=5$ (right) shows suitable features that persist to $n_{\rm I}\ge5$, reduce to approximately circular objects, and have a small distance between the centre of light and the maximum intensity position for $n_{\rm I}=5$. The centre of light of the most unstable feature $(15,106)$ of Fig.~\ref{fig:Image_2_loc} is marked with a white X to show the smallness of the difference between peak and centre of light.
  }
  \label{fig:Image_2_persistence}
\end{figure*}

In this parameter configuration, we extract all objects for all images in Fig.~\ref{fig:Image_bkg_CL0024} for $n_{\rm{I}}\ge3$ and plot their persistence diagrams to determine which of these objects are stable features that can be matched across all multiple images. 
First, the longer an object persists in the diagram over increasing thresholds, the higher the probability it is a suitable feature. 
Second, objects showing horizontal lines in the persistence diagrams are also more likely to be suitable features because their centre of light coordinates do not vary, for instance, due to a large amount of noise. 
Third, the smaller the distance between the centre of light position, calculated as the first moment of an object's brightness distribution, and the peak position of maximum intensity, the more likely it is that the object under analysis is a suitable and not very noisy feature. 
Fourth, we would expect star-forming regions to be of approximately circular shape which are distorted into ellipses to first order by the gravitational lensing effect. 
Assuming that these features are far enough away from the critical curve not to be extremely distorted in the shape of arcs (see \cite{bib:Wagner7} for a detailed analysis in galaxy-cluster scale gravitational lenses), the semi-major and semi-minor axes of the second order moment of an object's brightness distribution also contribute to evaluate the usefulness of an object.
Highly elongated arcs can be excluded to be star-forming regions and only included in the feature set if such arcs are clearly identified across several multiple images as a more complex feature. 

Fig.~\ref{fig:Image_2_persistence} shows the detected objects and their properties for image~2 of Fig.~\ref{fig:Image_bkg_CL0024} at $n_{\rm I}=3$ (left) and $n_{\rm I}=5$ (right). 
We note that the many very faint objects that do not correspond to suitable features do not persist to $n_{\rm I}=5$ and are therefore easily excluded by the first criterion of persistence. 
As for the second criterion, we already saw the mostly horizontal lines in the persistence diagrams of Fig.~\ref{fig:Image_2_loc}. 
The largest change in position of the centre of light for the most unstable feature $(15,106)$ of Fig.~\ref{fig:Image_2_loc} is found to be 2 pixels, which is comparable to the most unstable features in the other multiple images and thereby sets the range of tolerable non-zero slopes for the second criterion. 
Comparing the centre of light (magenta dots) and maximum intensity (blue dots) positions within each of the objects passing the first two criteria in Fig.~\ref{fig:Image_2_persistence}, we note that the extended background structure of the galaxy and the proximity of individual features to each other cause the two positions to have distances of the order of 2 pixels at $n_{\rm I}\le5$. 
These distances are reduced below one pixel for increasing $n_{\rm I}>5$.
For the neighbouring features which overlap at low $n_{\rm I}$, reducing the \emph{deblend\_cont} may reveal the additional features visible by eye in the wings of the ellipses at larger $n_{\rm I}$.
Yet, as noted before, this change in parameter value also increases the amount of spurious, persistent features consisting of bright, locally restricted noise. 
The axes ratios of the semi-minor to semi-major axis of the quadrupole moments are also tracked over the increasing thresholds and remained greater than or equal to 0.15, thus giving an order of magnitude to identify star-forming regions as contained in the five images in CL0024. 

In summary, running the default configuration of the \texttt{sep} extraction routine and switching \emph{clean} off, the detected objects are analysed for their persistence at least existing to $n_{\rm I}\le3$, stability of the centre of light position with only 2 pixel variations over all $n_{\rm I}$, the distance between the latter and the maximum intensity position of up to 1 pixel at the highest $n_{\rm I}$, and an only slightly elliptical quadrupole moment of the intensity distribution with axis ratios larger than 0.15.  
After determining the objects fulfiling these criteria, their properties at the highest $n_{\rm I}$ at which they can still be detected are used as stable features in the next steps\footnote{Our code is available at \url{https://github.com/joycelin1123/SEP-Automated-Feature-Detection}}. 


\subsection{Calibration of \texttt{ptmatch}}
\label{sec:calibration}

After the automated extraction of robust features using the criteria in Section~\ref{sec:feature_extraction}, we manually match the maximum number of corresponding features found across all images. 
For the example in Fig.~\ref{fig:Image_bkg_CL0024}, we find four features fulfiling the criteria of Section~\ref{sec:feature_extraction} that can be matched across all images.
They correspond to the features 1, 2, 4, and 5 identified by eye in \cite{bib:Wagner_cluster}.

These features can now be employed to run \texttt{ptmatch} and retrieve the local lens properties as detailed in Section~\ref{sec:theoretical_background}. 
In the idealised setting depicted in Section~\ref{sec:robust_features}, the peak positions as found by our \texttt{sep} pipeline would be inserted into \texttt{ptmatch}. 
However, as Section~\ref{sec:feature_extraction} indicated, an intensity maximum may be generated by noise, so that it is not the corresponding feature to the respective intensity peaks in other multiple images. 
Other lens reconstruction approaches which determine the `positions of multiple images' are not always clear about their precise choice of coordinates, it also depends on the image pre-processing. 
\cite{bib:Caminha2}, for instance, clearly use the peak position and thus assume all noise was eliminated. 
On the scale of star-forming regions as features, \cite{bib:Wagner7} showed that any physically caused offset between the peak and the centre of light, for instance due to higher-order lensing effects like flexion, is unlikely to occur and be resolved. 
This corroborates our third selection criterion for robust features of Section~\ref{sec:feature_extraction}. 

\begin{figure*}
  \centering
  \includegraphics[width=0.4\textwidth]{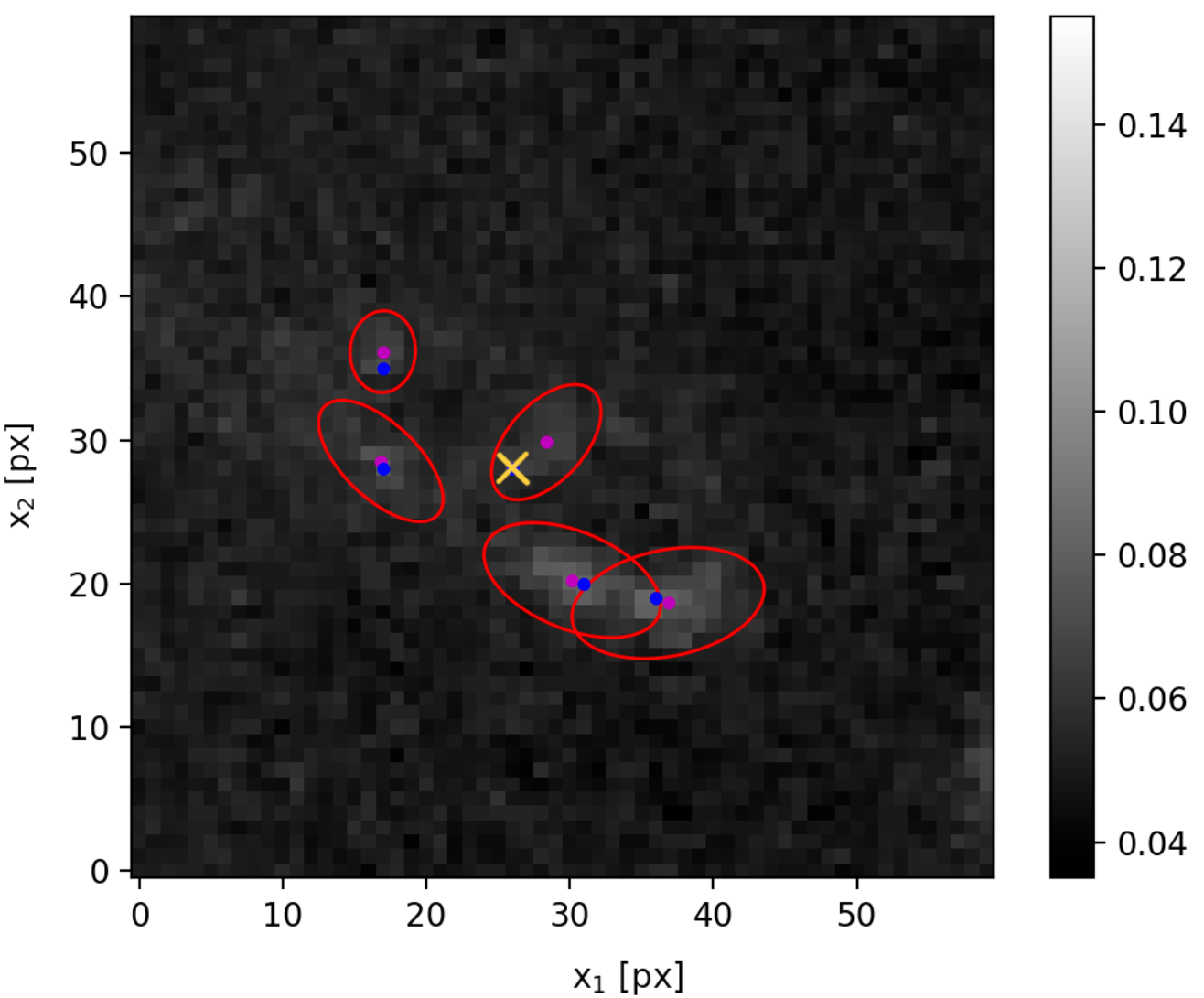}
  \hspace{0.1\textwidth}
  \includegraphics[width=0.4\textwidth]{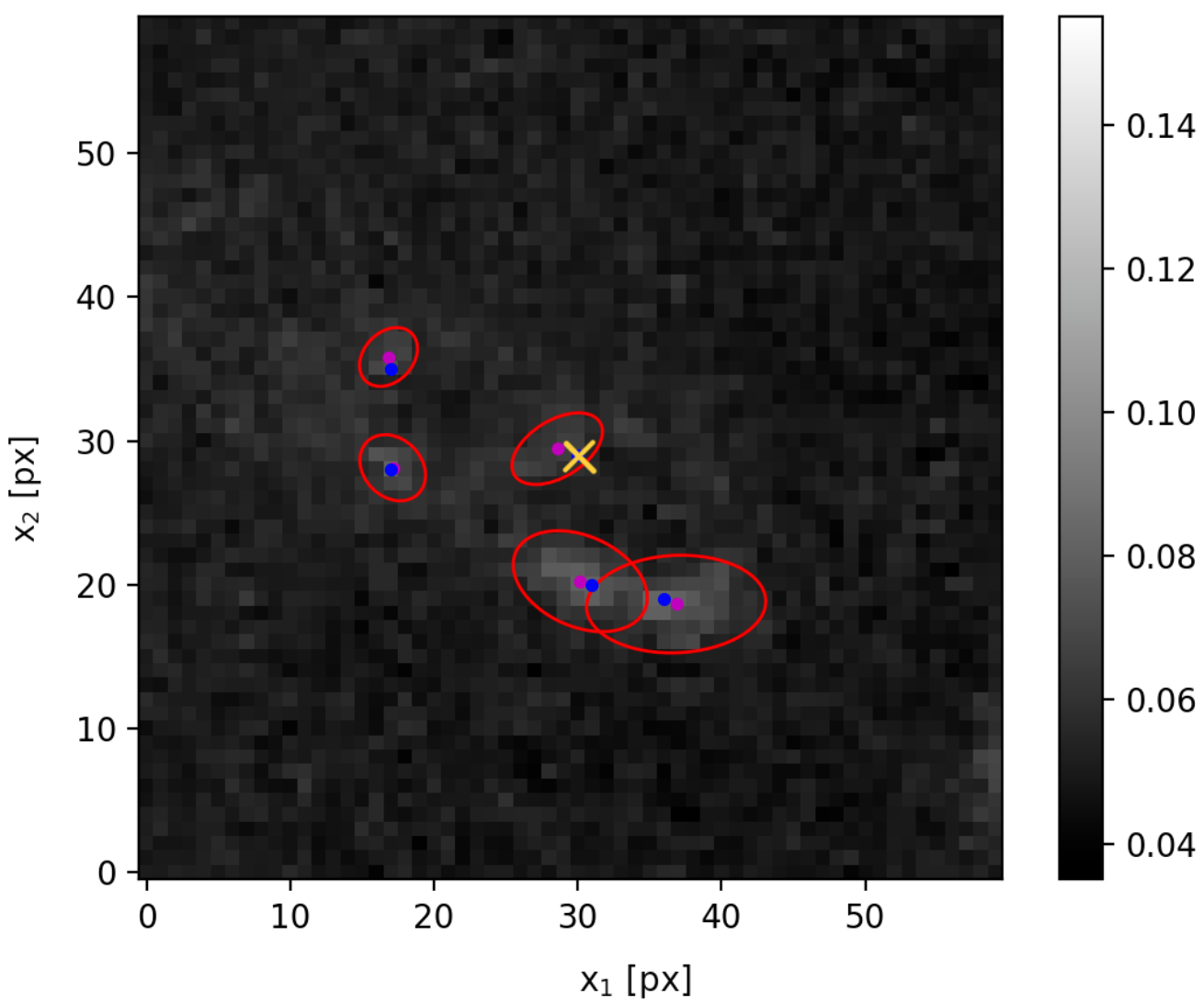}
  \caption{Identification of peak positions generated by noise for the example of image~5 in CL0024 observed in the F555W filter band, thresholded at $n_{\rm I}=4.0$ (left) and $n_{\rm I}=4.6$ (right). The central object's peak position (yellow cross) at the last detectable threshold with $n_{\rm I}=4$ is several pixel away from the centre of light position (magenta dot). Increasing the threshold to $n_{\rm I}=4.6$, the peak position shifts closer to the centre of light. Thus the peak is no genuine maximum intensity value but an artefact due to the low signal-to-noise environment.
  }
  \label{fig:Image_5_peak}
\end{figure*}

The persistence analysis introduced in Section~\ref{sec:feature_extraction} allows to evaluate the quality of the chosen features to be matched. 
Tracking the peak position with respect to the centre of light position in each object over increasing intensity thresholds, peaks created by noise and peaks representing genuine intensity maxima can be distinguished. 
A second example, in addition to the one in Fig.~\ref{fig:Image_2_persistence}, is the object detection in the central image~5 in CL0024 as observed in the F555W filter band shown in Fig.~\ref{fig:Image_5_peak}.
This image is located in a low signal-to-noise environment.
Therefore, at the highest threshold, $n_{\rm I}=4$, that identifies the object in the centre of Fig.~\ref{fig:Image_5_peak} (left), the peak position (yellow cross) is still 3.2 pixel away from the centre of light position (magenta dot). 
To investigate the cause of the distance, we manually increase the threshold to $n_{\rm{I}}=4.6$, at which the object is still detected. 
As can be read off the comparison between Fig.~\ref{fig:Image_5_peak} (left) and (right), the peak position does not shift closer to the centre of light along the connection between the peak and the centre of light, but completely changes its position with respect to the centre of light. 
Thus, this unstable peak should not be employed as a feature to be used in \texttt{ptmatch}. 
According to the feature selection criteria of Section~\ref{sec:feature_extraction}, the object is discarded due to the distance between the peak and the centre of light exceeding one pixel. 

\begin{figure}
  \centering
  \includegraphics[width=0.49\textwidth]{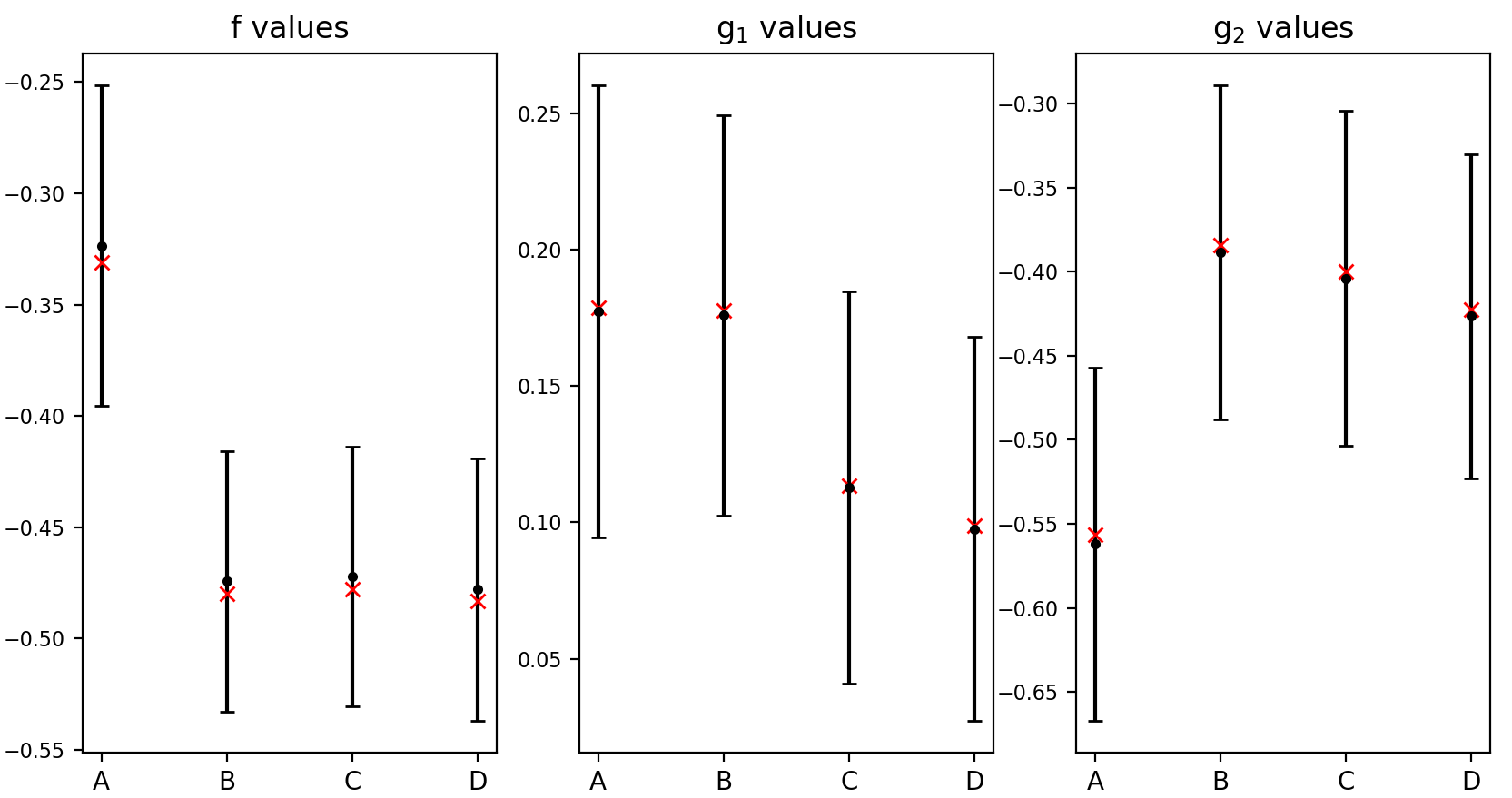}
  \vspace{-3ex}
  \caption{Impact of different feature position choices on inferred local lens properties $f$, $g_1$, and $g_2$ for image~5 of Fig.~\ref{fig:Image_5_peak}. 
  A: results using peak points at $n_{\rm I}=4$ ($\chi^{2}_{\rm red}$ = 1.013), 
  B: results using peak points at $n_{\rm I}=4.6$ ($\chi^{2}_{\rm red}$ = 0.459),
  C: results using peak points at $n_{\rm I}=4$ for all features but the noisy one which is replaced by the centre of light ($\chi^{2}_{\rm red}$ = 0.623),
  D: results using centre of light positions at $n_{\rm I}=4$ for all features ($\chi^{2}_{\rm red}$ = 0.478). Plots show the mean local lens property (black mark) with its 68\% confidence bounds (error bars) and the most likely local lens property (red mark) as obtained from \texttt{ptmatch}.}
  \label{fig:Image_5_fg}
\end{figure}

Yet, cases may arise in which such objects are erroneously employed as features or have to be considered for the lack of qualitatively better ones. 
To test the impact of these inferior features on the \texttt{ptmatch} results, Fig.~\ref{fig:Image_5_fg} shows the local lens properties determined for image~5 of Fig.~\ref{fig:Image_5_peak} (further details on this and other filter bands are detailed in Section~\ref{sec:cl0024}).  
The measurement uncertainties for the positions, $\sigma_1$ and $\sigma_2$ for the $x_1$- and $x_2$- coordinates respectively, were set to one pixel in all cases. 
Then, \texttt{ptmatch} was run on all multiple images for the F555W filter band to determine the local lens properties of Eqs.~\ref{eq:f} and \ref{eq:g}. 
Leaving all feature coordinates in all other multiple images at the stable peak positions, different choices of positions as detailed in Fig.~\ref{fig:Image_5_fg} are used for the features in image~5\footnote{The uppermost small object around $(19,37)$ in Fig.~\ref{fig:Image_5_peak} is discarded from the set of features for the lack of corresponding ones in other images.}.  
Comparing the $f$ and $\boldsymbol{g}$ values for the four different choices, employing the noisy peak position leads to deviations in the inferred local lens properties beyond consistencies within the 68\% confidence bounds marked by the error bars. 
The comparison of the two position choices shows that local lens properties as inferred from positions that vary on the scale of one pixel are consistent with each other and that the $\chi^2_{\rm red}$ is not the only measure to be monitored when evaluating the overall quality of fit, as it does not indicate whether the chosen position for a feature is physically meaningful.

Having investigated the impact of the coordinate choices on the inferred local lens properties, we continue to apply our pipeline for automated feature extraction based on \texttt{sep} and subsequent inference of local lens properties by \texttt{ptmatch} to multiple wavebands for CL0024 and the galaxy-cluster containing Hamilton's Object. 
Unless noted otherwise, we apply the feature selection criteria of Section~\ref{sec:feature_extraction}, choose the stable peak positions, and run \texttt{ptmatch} in its standard configuration with measurement uncertainties of $\sigma_1=\sigma_2=1~\mbox{px}$, motivated by the typical distances between the peak and the centre of light position. 

\section{Applications}
\label{sec:applications}

We now apply our robust feature extraction pipeline to the two multiple-image configurations previously analysed in \cite{bib:Wagner_cluster} and \cite{bib:Griffiths} and extend the manual feature extraction to multiple wavelengths in the case of CL0024 (detailed in Section~\ref{sec:cl0024}) and to a feature extraction from individual filter bands (instead of a joint picture) for the cluster containing Hamilton's Object (detailed in Section~\ref{sec:hamiltons_object}). 
Subsequently, we run \texttt{ptmatch} on the manually matched features and obtain the local lensing properties, Eqs.~\ref{eq:f}, \ref{eq:g}, and \ref{eq:j} for all filter bands. 
Tables~\ref{tab:CL0024_summary} and \ref{tab:Hamilton_summary} summarise all local lens properties obtained for both cases. 

\subsection{Five-image configuration in CL0024}
\label{sec:cl0024}
\begin{table}
		\caption{Abbreviating labels used to in Figs.~\ref{fig:cl0024_image_1}, \ref{fig:cl0024_image_2}, \ref{fig:cl0024_image_3}, \ref{fig:cl0024_image_4}, and \ref{fig:cl0024_image_5} to denote the filter band and extraction method used to infer the local lens properties. The four features used in most cases are features~1, 2, 4, and 5 in \protect\cite{bib:Wagner_cluster}. The last column lists the $\chi_{\rm{red}}$ values obtained by \texttt{ptmatch}.}
		\label{tab:CL0024_labels_table}		
		\begin{center}
			\begin{tabular}{c|l|r}
				\hline
				Label & Waveband (features, extraction) & $\chi^2_{\rm red}$ \\
				\hline
				A &  F475W  (4 features by eye) & 1.35 \\
				B &  F475W (6 features by eye) & 1.73 \\
				C & F435W (4 features by \texttt{sep}) & 0.75 \\
				D & F475W (4 features by \texttt{sep}) & 0.15 \\
				E & F555W (4 features by \texttt{sep}) & 0.62 \\
				F & F625W (4 features by \texttt{sep}) & 0.69 \\
				G & F775W (4 features by \texttt{sep}) & 1.41 \\
			    H & F850LP (4 features by \texttt{sep}) & 2.70 \\
				\hline                                  
			\end{tabular}
		\end{center}
\end{table}

The first example is the five-image configuration of a blue spiral galaxy at $z_\mathrm{s}=1.675$ behind the galaxy cluster CL0024+1654, abbreviated as CL0024, at $z_\mathrm{d}=0.39$, which we already investigated in \cite{bib:Wagner_cluster} using the approach of Section~\ref{sec:summary_formalism}. 
Here, we re-analyse this example using the automatically extracted features to demonstrate that an objective feature extraction yields similar local lens properties as features identified by eye. 
The standardised feature extraction also allows us to compare the local lens properties obtained in several wavelength filter bands to investigate the wavelength independence of strong gravitational lensing. 

\begin{figure}
  \centering
  \includegraphics[width=0.49\textwidth]{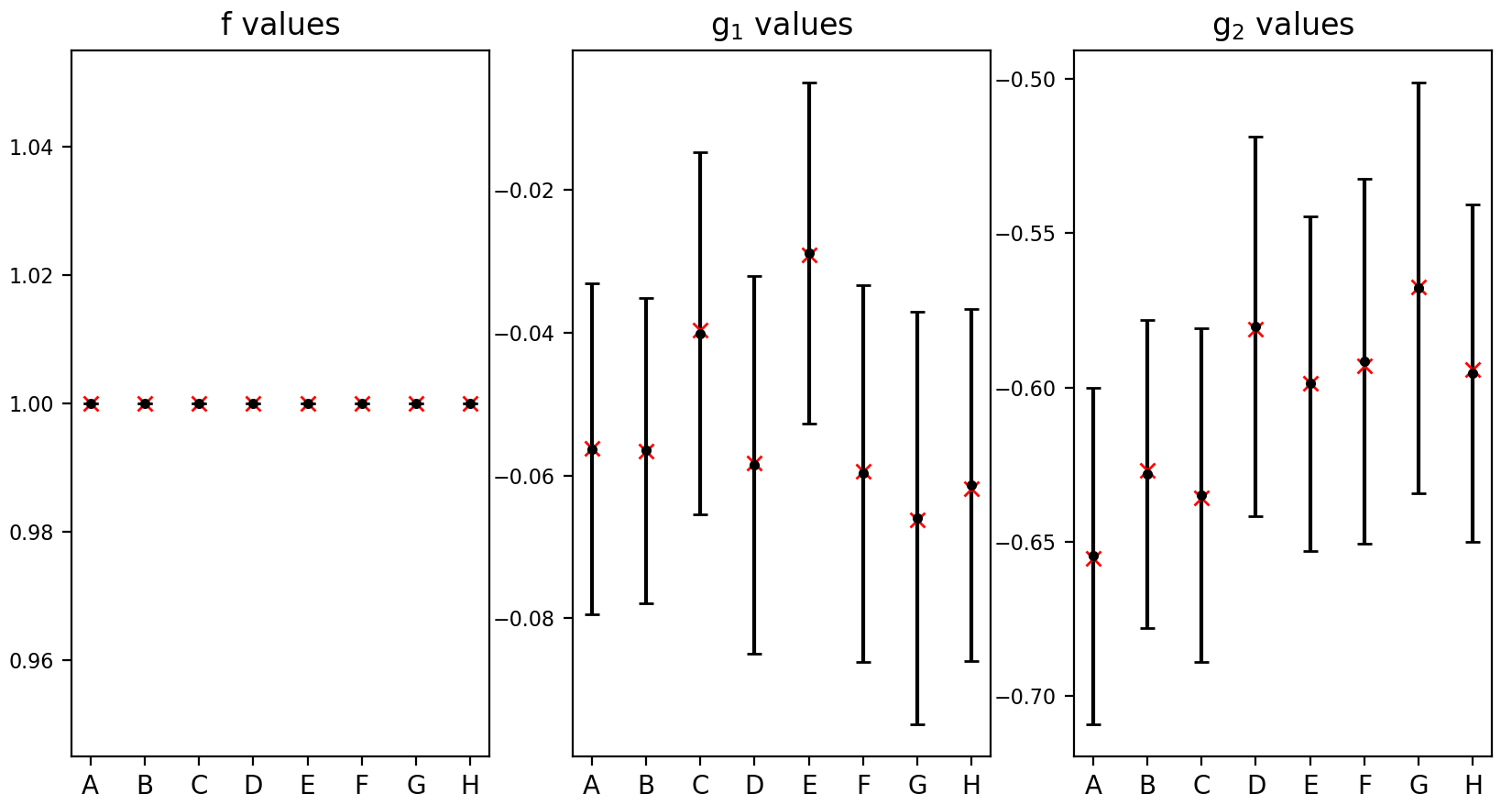}
    \vspace{-3ex}
  \caption{Local lens properties for image~1 in the CL0024 over six wavebands, abbr.~A--H as in Table~\ref{tab:CL0024_labels_table}, mean (black dot), 68\% confidence intervals (error bars), and most likely value (red cross).}
  \label{fig:cl0024_image_1}
\end{figure}

\begin{figure}
	\centering
	\includegraphics[width=0.49\textwidth]{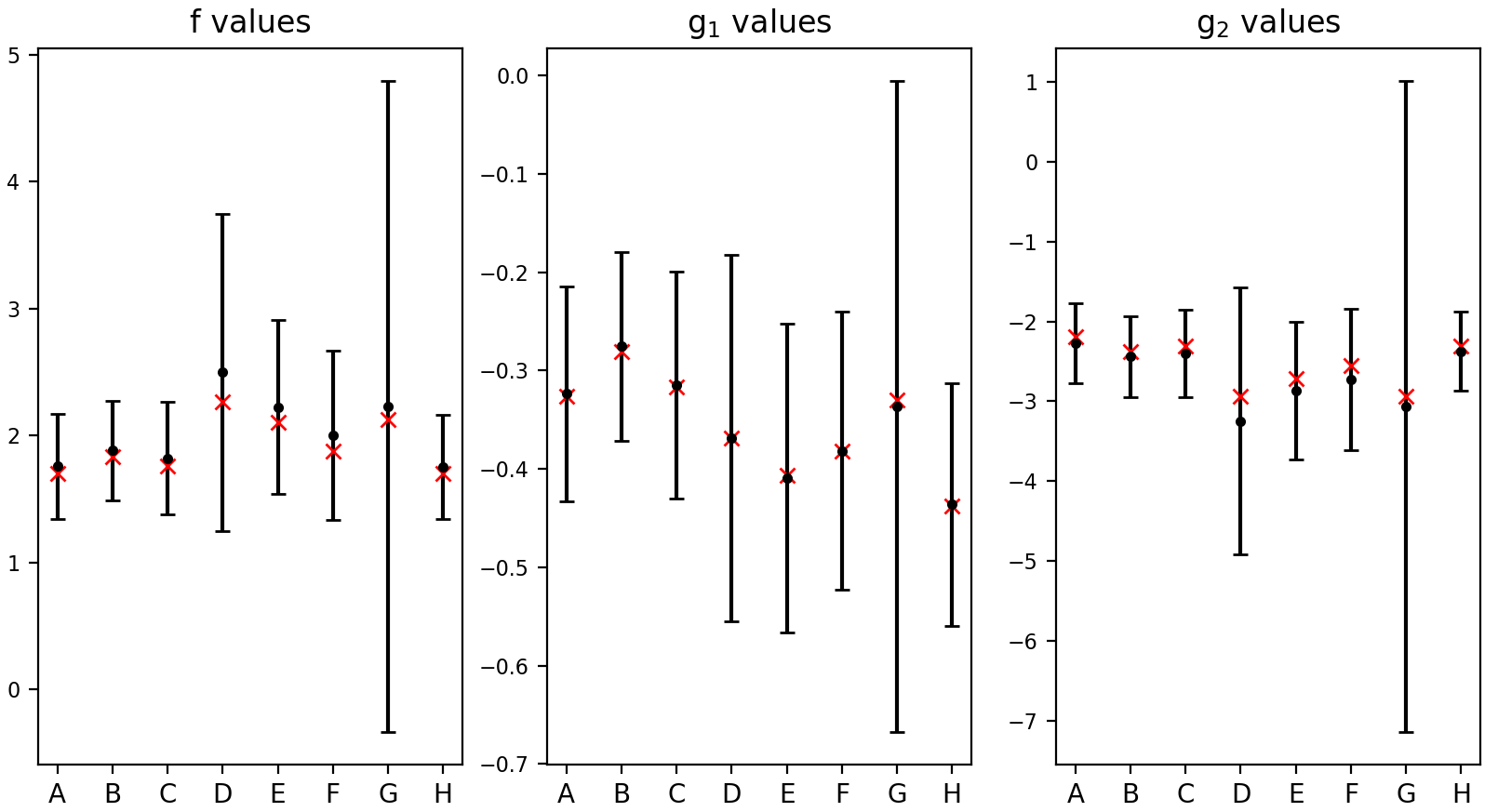}
	  \vspace{-3ex}
	\caption{Same as Fig.~\ref{fig:cl0024_image_1} for image~2 in CL0024.}
	\label{fig:cl0024_image_2}
\end{figure}

\begin{figure}
  \centering
  \includegraphics[width=0.49\textwidth]{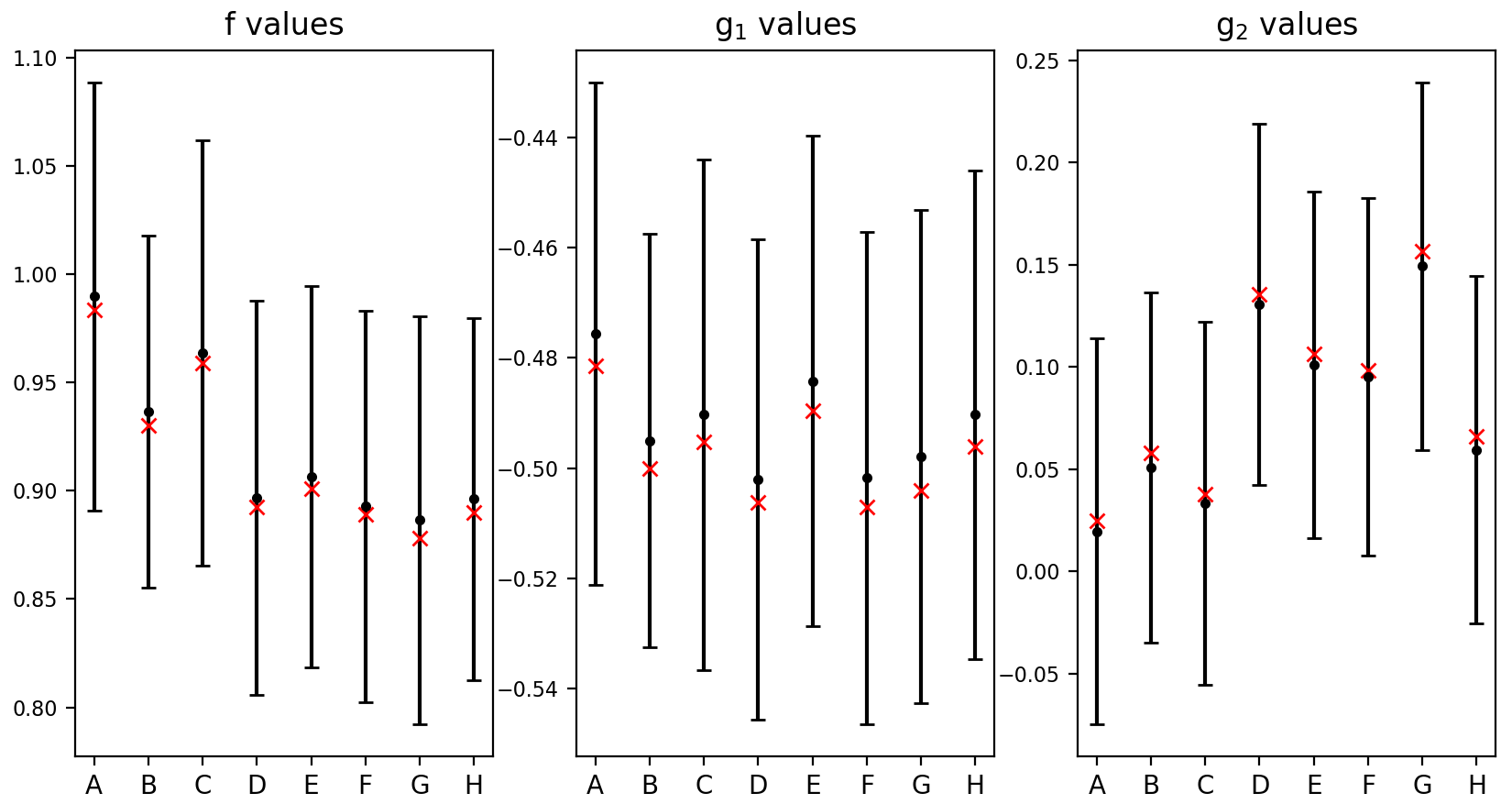}
  \vspace{-3ex}
  \caption{Same as Fig.~\ref{fig:cl0024_image_1} for image~3 in CL0024.}
  \label{fig:cl0024_image_3}
\end{figure}

\begin{figure}
  \centering
  \includegraphics[width=0.49\textwidth]{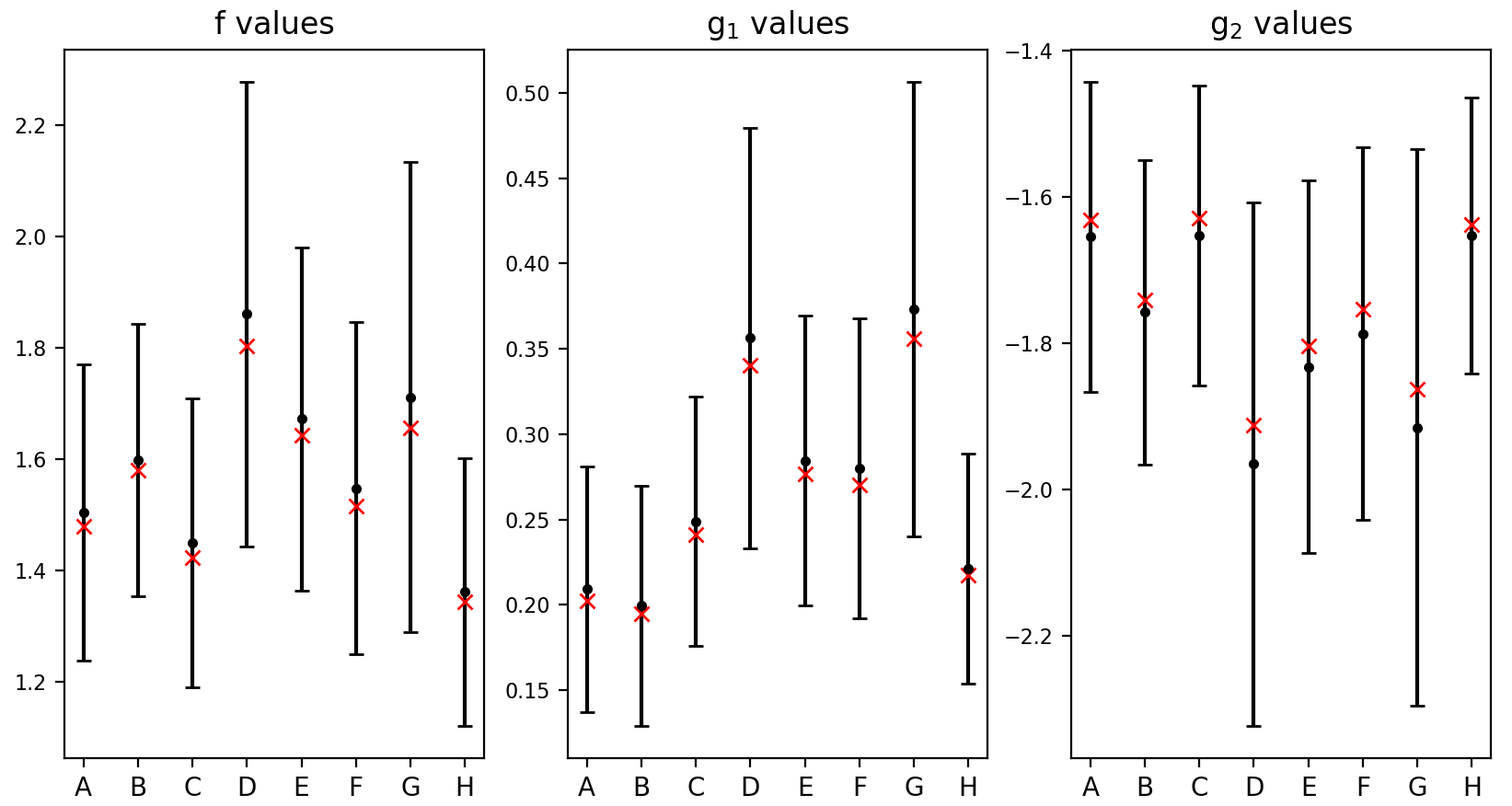}
    \vspace{-3ex}
  \caption{Same as Fig.~\ref{fig:cl0024_image_1} for image~4 in CL0024.}
  \label{fig:cl0024_image_4}
\end{figure}

\begin{figure}
  \centering
  \includegraphics[width=0.49\textwidth]{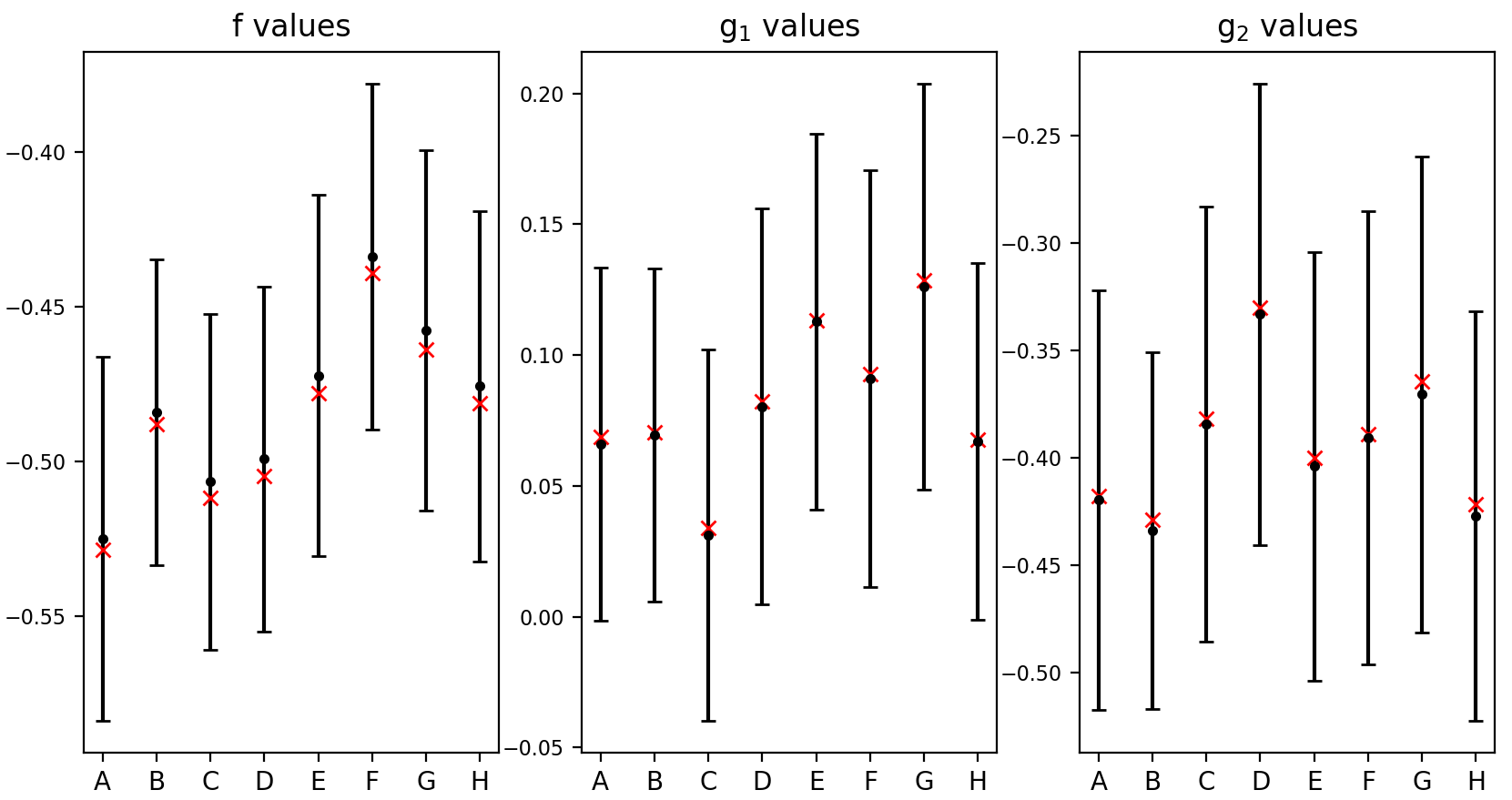}
    \vspace{-3ex}
  \caption{Same as Fig.~\ref{fig:cl0024_image_1} for image~5 in CL0024.}
  \label{fig:cl0024_image_5}
\end{figure}

The observations used here were made by HST in November 2004 (PI: Ford) using filters F435W, F475W, F555W, F625W, F775W and F850LP with exposure times 6435s, 5072s, 5072s, 7756s, 5072s, and 8164s, respectively. 
Each waveband has the same spatial resolution, but different signal-to-noise ratios, as the background estimation reveals.
The feature extraction as outlined in Section~\ref{sec:robust_feature_extraction} is performed for all filter bands. 
A manual cross-check confirms that it leads to a stable detection of features~1, 2, 4, and 5, the star-forming regions used as manually identified features, in \cite{bib:Wagner_cluster}. 
Out of the 24 required feature identifications, the only noisy peak is the one in image~5 in F555W, shown in Fig.~\ref{fig:Image_5_peak}, which is replaced by the centre of light, as detailed in Section~\ref{sec:calibration}, because the feature is vital to span a larger area (see \cite{bib:Wagner_cluster} on the impact of the image area on the precision of the local lens properties).
Our manual follow-up analysis showed that the true stable peak lies close to the centre of light and therefore, the local lens properties as obtained from the automated feature extraction for the centre of light are in agreement with the one from the true peak within the confidence bounds in this case.
Features~3 and 6 of \cite{bib:Wagner_cluster} were found in larger images with a higher signal-to-noise ratio such as images~1 and 4, but could not reliably be detected in images with lower signal to noise such as image~5.

In Figs.~\ref{fig:cl0024_image_1}--\ref{fig:cl0024_image_5}, we compare all local lens properties for each image over the six filter bands and with the local lens properties that are obtained for the same four manually identified features and all six manually identified features of \cite{bib:Wagner_cluster}. 
To shorten notations, Table~\ref{tab:CL0024_labels_table} lists the abbreviations used as labels on the abscissa of the plots and the $\chi_{\rm{red}}^2$ values for the best-fit obtained by \texttt{ptmatch}. 

Comparing the $\chi_{\rm{red}}^2$ values between the manual feature identification and the automated one, we find that, except for F775W and F850LP, the matching is improved and the $\chi_{\rm{red}}^2$ values reduced from a biased fitting to over-fitting. 
The F475W filter band shows the lowest $\chi_{\rm{red}}^2$ values, potentially due to its role as calibration example.
Yet, Figs.~\ref{fig:cl0024_image_1}--\ref{fig:cl0024_image_5} show no signs of specialty with respect to other filter bands.
Confidence bounds for image~2 and image~4 are even the second largest for F475W, only surpassed by those in F775W.
Section~\ref{sec:hamiltons_object} will show that similarly low $\chi_{\rm red}^2$ are obtained in a different setting as well (see Table~\ref{tab:HO_labels_table}) to support the claim that F475W yields typical results. 
Thus, the feature extraction generalises as expected to other filter bands. 
Yet, the increased $\chi_{\rm{red}}^2$ values for F775W and F850LP and the increased confidence bounds particularly observed for F775W show that the automated detection reaches its limits for those filter bands.
The emission profile of the blue background galaxy fades into the background light for these wavelengths. 
This effect is clearly visible in Fig.~\ref{fig:cl0024_image_2} for image~2, which requires a very precise feature identification because it is located close to an isocontour of $\kappa=1$, as detailed in \cite{bib:Wagner_cluster}. 

\begin{figure*}
	\centering
	\makebox[\linewidth]{
		\includegraphics[width=\textwidth]{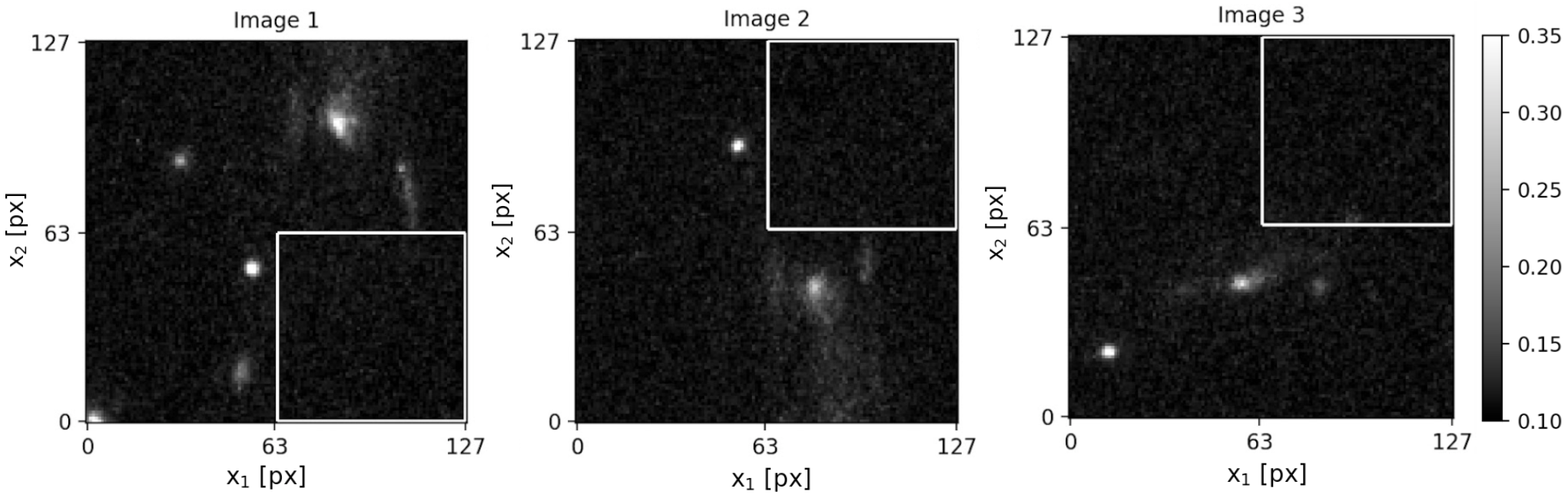}
	}
	\caption{Original data used in Section~\ref{sec:hamiltons_object} for feature detection: Image~1--3 in the cluster-scale lens J2230 in the HST ACS/WFC F606W filter band, analogous to Fig.~\ref{fig:Image_bkg_CL0024} for the five-image configuration in CL0024. Image~1 and 2 are a fold configuration called Hamilton's Object.}
	\label{fig:HO_all}
\end{figure*}

As also observed from Fig.~\ref{fig:cl0024_image_1}--\ref{fig:cl0024_image_5}, all local lens properties are in agreement within the 68\% confidence bounds. 
For further details, Table~\ref{tab:CL0024_summary} contains all values.
Variations in the value and the size of the confidence bounds can be observed, most significantly for images~2 and 4 in F475W and F775W as noted above. 
Altogether no significant disagreement is found within the current precision.
We can thus conclude that no deviations from a wavelength independence of strong lensing can be found in this multiple-image configuration.

As a cross-check, we determine the magnification ratios of Eq.~\ref{eq:j}. 
The results are plotted in App.~\ref{app:cl0024_magnification_ratios} for completeness. 
As expected, these graphs also show a high degree of agreement and thus yield the same conclusion. 
Compared to the $f$ and $\boldsymbol{g}$ values, their confidence bounds are smaller (due to the implementation of \texttt{ptmatch}), such that they are even a better test for constant local lens properties over all wavelengths. 
To investigate potential biases of the local lens properties, for instance, by dust extinction or micro-lensing, a comparison to the measured flux ratios is required. 
This can be easily performed for multiple images like Hamilton's Object, detailed below in Section~\ref{sec:hamiltons_object}.
Its main flux is contained in a central bright feature and can thus be determined by a standard flux measurement implemented in \texttt{SExtractor} or \texttt{sep}. 
For this image configuration in CL0024, however, such a flux measurement is hard to set up because it should contain the flux collected from the area spanned by all features.
This flux cannot be calculated with the standard astrometry and photometry programs because they only focus on individual objects like the star-forming regions but not arbitrarily shaped areas. 
So we leave such a comparison, which is beyond the scope of this work, for future investigations. 

\subsection{Three-image configuration with Hamilton's Object}
\label{sec:hamiltons_object}

\begin{table}
		\caption{Abbreviating labels used in Figs.~\ref{fig:4t_4b}, \ref{fig:ptmatch_Image_1}, \ref{fig:ptmatch_Image_2}, \ref{fig:ptmatch_Image_3}, \ref{fig:HO_j} to denote the filter band and extraction method used to infer the local lens properties. CI denotes the combination of F110W, F606W$_{\rm d}$, and F815W$_{\rm d}$ to a coloured image, the subscript d indicates a downsized version, details see text. The four features used in most cases are features~1, 3, 4, and 7 in \protect\cite{bib:Griffiths}. The last column lists the $\chi_{\rm{red}}$ values obtained by \texttt{ptmatch}.}
		\label{tab:HO_labels_table}		
		\begin{center}
			\begin{tabular}{c|l|r}
				\hline
				Label & Waveband (features, extraction) & $\chi^2_{\rm {red}}$ \\
				\hline
				A & CI (4 features by eye)  & 0.06 \\
				B & CI (7 features by eye)  & 1.11 \\
				C & F110W (4 features by \texttt{sep}) & 0.87 \\
				D & F606W$_{\rm d}$ (4 features by \texttt{sep}) & 0.12 \\
				E & F814W$_{\rm d}$ (4 features by \texttt{sep}) & 0.73 \\
				F & F606W (4 features by \texttt{sep}) & 0.12 \\		
				G & F814W (4 features by \texttt{sep}) & 0.42 \\	
				H & CI (4 features by \texttt{sep}) & 0.22 \\
				\hline                              
			\end{tabular}
		\end{center}
\end{table}

Our second example is the triple-image configuration containing Hamilton's object as first analysed in \cite{bib:Griffiths}.
The background source is a star-forming galaxy at $z_\mathrm{s}=0.8200$ multiply-imaged by the galaxy cluster RM J223013.1-080853.1, J2230 for short, at $z_\mathrm{d}=0.526$. 
Up to our knowledge, Hamilton's Object is the first fold configuration with almost perfectly aligned images orthogonal to their straddling critical curve which also has an extended counter-image clearly showing the same substructure, as spectroscopically identified in \cite{bib:Griffiths}. 
Fig.~\ref{fig:HO_all} gives an overview of the three images. 

For this case, your focus lies in first determining robust features for each individual filter band from the automated pipeline outlined in Section~\ref{sec:robust_feature_extraction}, which could not be achieved manually. 
To identify matching features by eye, we had to create a coloured image as a combination of the F606W, F814W, and F110W filter bands.
Assembling this coloured image was an image processing task on its own, as detailed in \cite{bib:Griffiths}. 
The F606W and F814W filter bands with a resolution of $1.4~\times~10^{-5}$ degrees per pixel observed by ACS/WFC had to be downsized and aligned with the F110W filter band with a resolution of $2.5~\times~10^{-5}$ degrees per pixel observed by WFC3/IR. 
Thus, this triple-image configuration can also be used to study the robustness of our feature extraction and the inferred local lens properties for the same data at different resolutions. 
Our second goal here is therefore to compare the local lens properties inferred for F606W and F814W and their downsized versions, abbreviated as F606W$_{\rm d}$ and F814W$_{\rm d}$, respectively. 

The three images denoted as A, B, and C in \cite{bib:Griffiths} will be referred to as images~2, 1, and 3 respectively because we now use image~1 (image~B in \cite{bib:Griffiths}) as a reference image for \texttt{ptmatch} to obtain more robust confidence bounds for the local lens properties and $\chi^2_{\rm {red}}$ closer to 1. 
The observations used in this paper of Hamilton's Object were made by HST with the filter bands F606W and F814W taken in 2015 and the band F110W taken in 2016 (PI: Ebeling for both). 
They had exposure times of 1200s, 1200s, and 705.88s, respectively. 

Running our feature extraction pipeline as detailed in Section~\ref{sec:robust_feature_extraction} for all filter-band configurations listed in Table~\ref{tab:HO_labels_table}\footnote{The automated detection of features in the coloured image is performed on its conversion into a grey-valued image by means of the Python Imaging Library (PIL) function `convert(''L'')'.}, features~1, 3, 4, and 7 of \cite{bib:Griffiths} are robustly identified without any noisy peaks. 
Hence, the automated feature extraction succeeds in identifying features across different filter bands with different signal-to-noise ratios and different resolutions, which was not possible by eye. 

\begin{figure}
	\centering
	\includegraphics[width=0.4\textwidth]{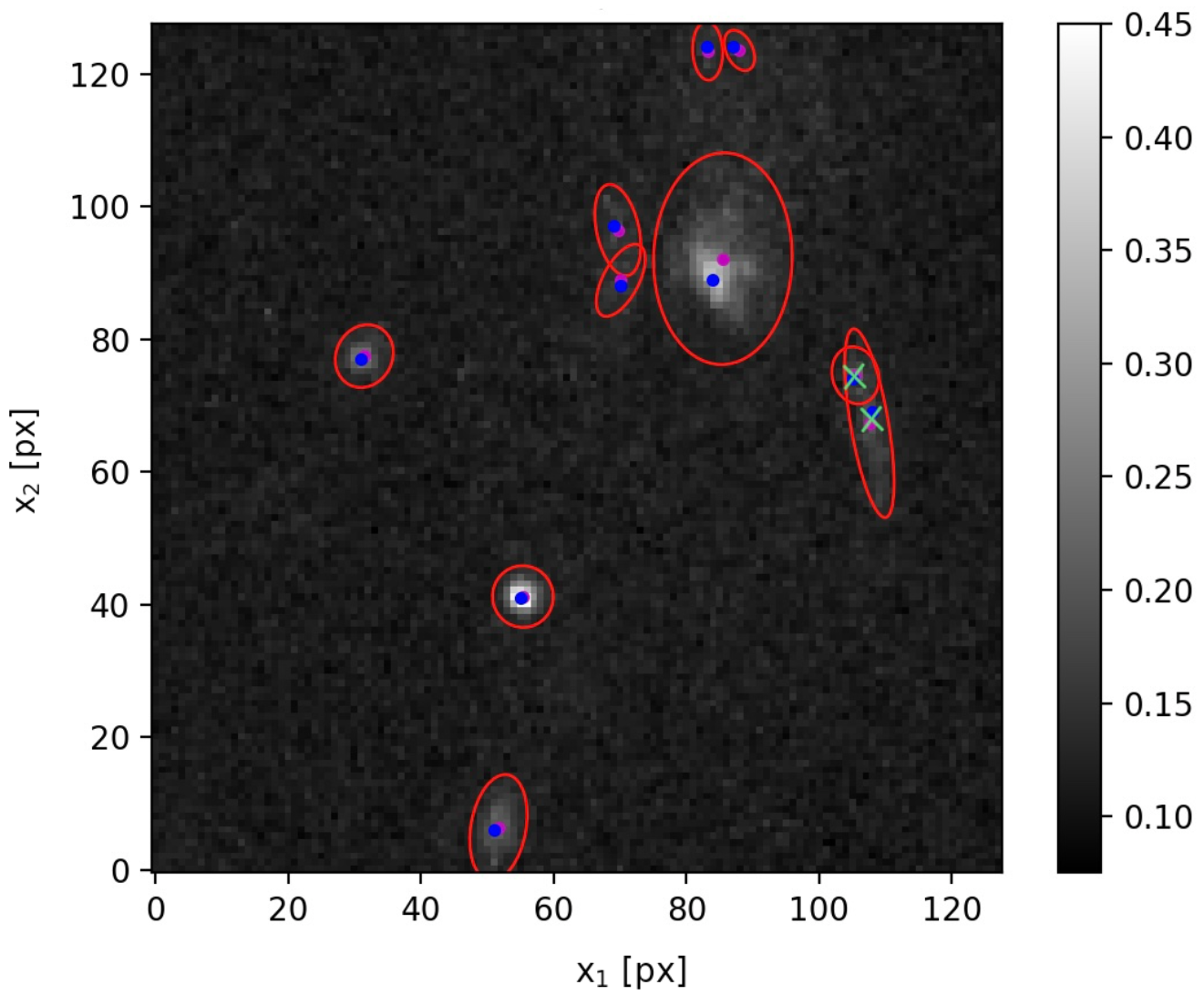}
	\caption{Detection of an additional robust feature in image~1 in J2230. Matching feature~4 in image~1, two choices (green crosses) with peak-to-peak separation of 6 pixels show up. The plot shows the detected objects in images~1 in F606W waveband at a $n_{\rm I}$ = 3, notation as in Fig.~\ref{fig:Image1_sig5}.
	}
	\label{fig:feature_split}
\end{figure}

Further investigating all identified features, we note an interesting detection in the F606W, F814W, their downsized versions, and the coloured image as shown in Fig.~\ref{fig:feature_split} in the F606W filter band. 
While all other images clearly only have one object detection that can be considered as feature~4, two possible objects show up in the above mentioned filter-band configurations. 
As the two almost vertically aligned features, called 4t and 4b (top feature and bottom feature, respectively), are not visible in the F110W filter band image, one may interpret them as a split of feature~4 into two parts due to the increased resolution in the F606W and F814W filter bands. 
Yet, the downsized versions also show feature~4t and 4b, so that this explanation can be excluded. 
With the same argument, it can be rejected that image~1 being the largest image of all images may show more details of the background source because it is subject to a stronger gravitational lensing effect. 
Thus, two interpretations remain.
The second additional feature can have a narrow emission profile restricting its visibility to the optical wavelengths.
Alternatively, given that the two observations happened about a year apart, the second feature can be a transient micro-lensing event similar to the ones made by \cite{bib:Kelly2} and \cite{bib:Kaurov}, explained by a two-scale lens modelling of the global galaxy-cluster and the local micro-lensing scale in \cite{bib:Diego}. 

\begin{figure}
	\centering
	\includegraphics[width=0.49\textwidth]{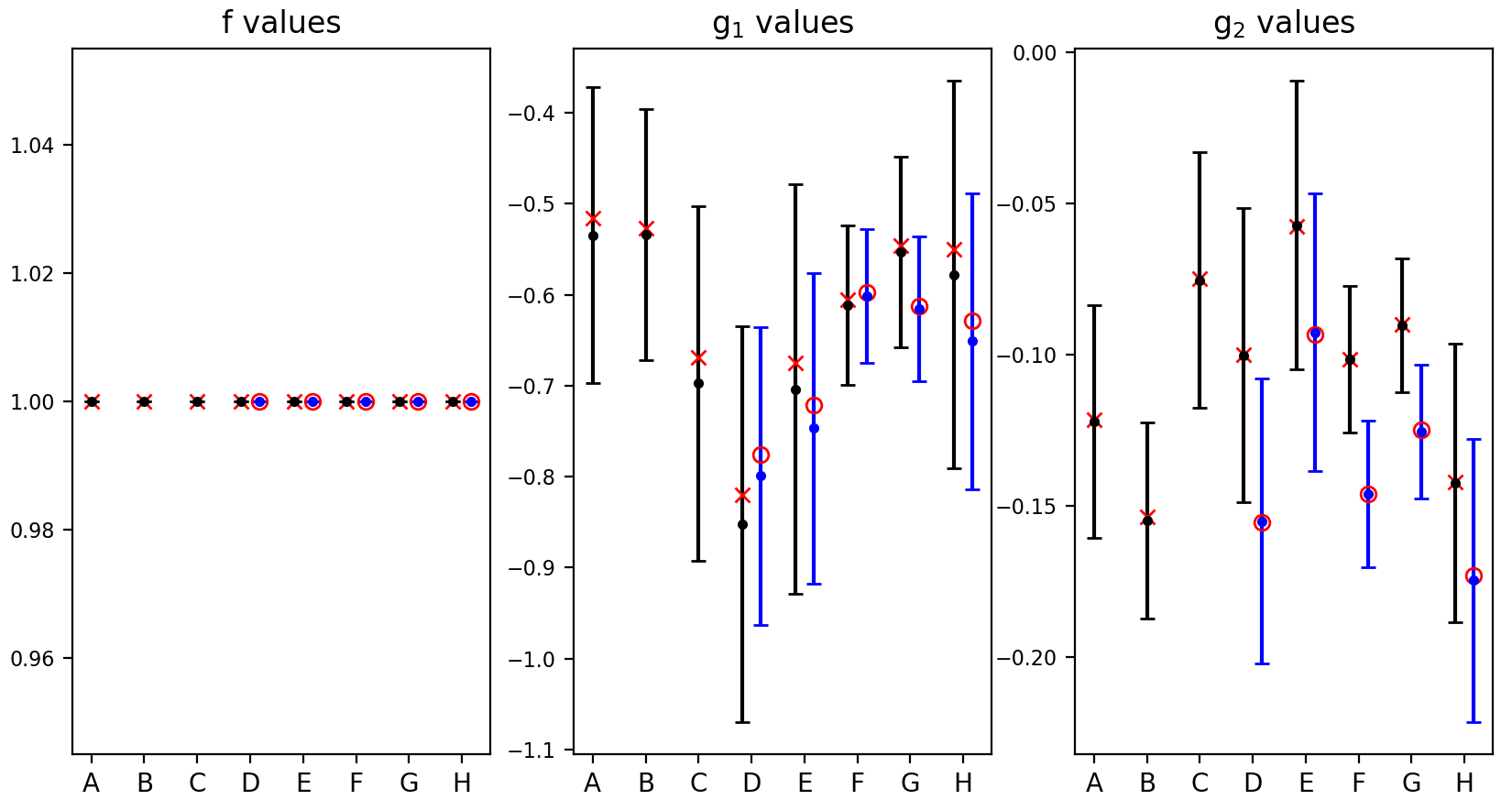}
	\vspace{-3ex}
	\caption{Local lens properties for image~1 in J2230 for all filter-band configurations in Table~\ref{tab:HO_labels_table} using feature~4t (black lines) or 4b (blue lines) as feature~4 for image~1 (see Fig.~\ref{fig:feature_split}). Notation analogous to Fig.~\ref{fig:cl0024_image_1}.
	}
	\label{fig:4t_4b}
\end{figure}

Comparing intensities and relative positions, feature~4t is considered the best matching one and will be used in all following plots. 
Yet, assuming we use feature~4b instead, we can perform a similar test for robustness of \texttt{ptmatch} as for image~5 in CL0024 detailed in Section~\ref{sec:calibration}. 
Fig.~\ref{fig:4t_4b} shows a comparison of all local lens properties in image~1 obtained over all filter-band configurations of Table~\ref{tab:HO_labels_table} when using feature~4t (black lines) or 4b (blue lines) as feature~4. 
The $\chi^2_{\rm red}$ values for the \texttt{ptmatch} fits are of the same order of magnitude for both configurations.
The plots for images~2 and 3 look similar to the $g_1$-value plot in Fig.~\ref{fig:4t_4b} and are thus insensitive to the feature exchange. 
They are plotted in Appendix~\ref{app:4t_4b}. 
In contrast, the $g_2$ values in image~1 show a shift to smaller $g_2$ values, which is most prominent in the F606W and F814W filter bands. 
The unchanged values in all local lens properties except for the shift in the $g_2$ value of image~1 show that \texttt{ptmatch} only changes those local lens properties that are directly affected by a change in the relevant feature coordinates in the respective image. 
Larger $|g_2|$ values are expected because feature~4b is farther away from the bulge of other features and thus contributes to increase the amplitude of the shear in $g_2$ direction.
Comparing these results to the ones obtained by eye with four and seven features, there is a trend for the seven-feature configuration to agree with the local lens properties obtained using feature~4b, indicating that the remaining features also favour larger $|g_2|$ values.
In addition, the error bars are slightly smaller when using feature~4b (see also the plots in Appendix~\ref{app:4t_4b} that show less stable local lens properties for the downsized version of F606W and the colour image). 
Hence, the correct matching cannot be determined with certainty. 

\begin{figure}
	\centering
	\includegraphics[width=0.49\textwidth]{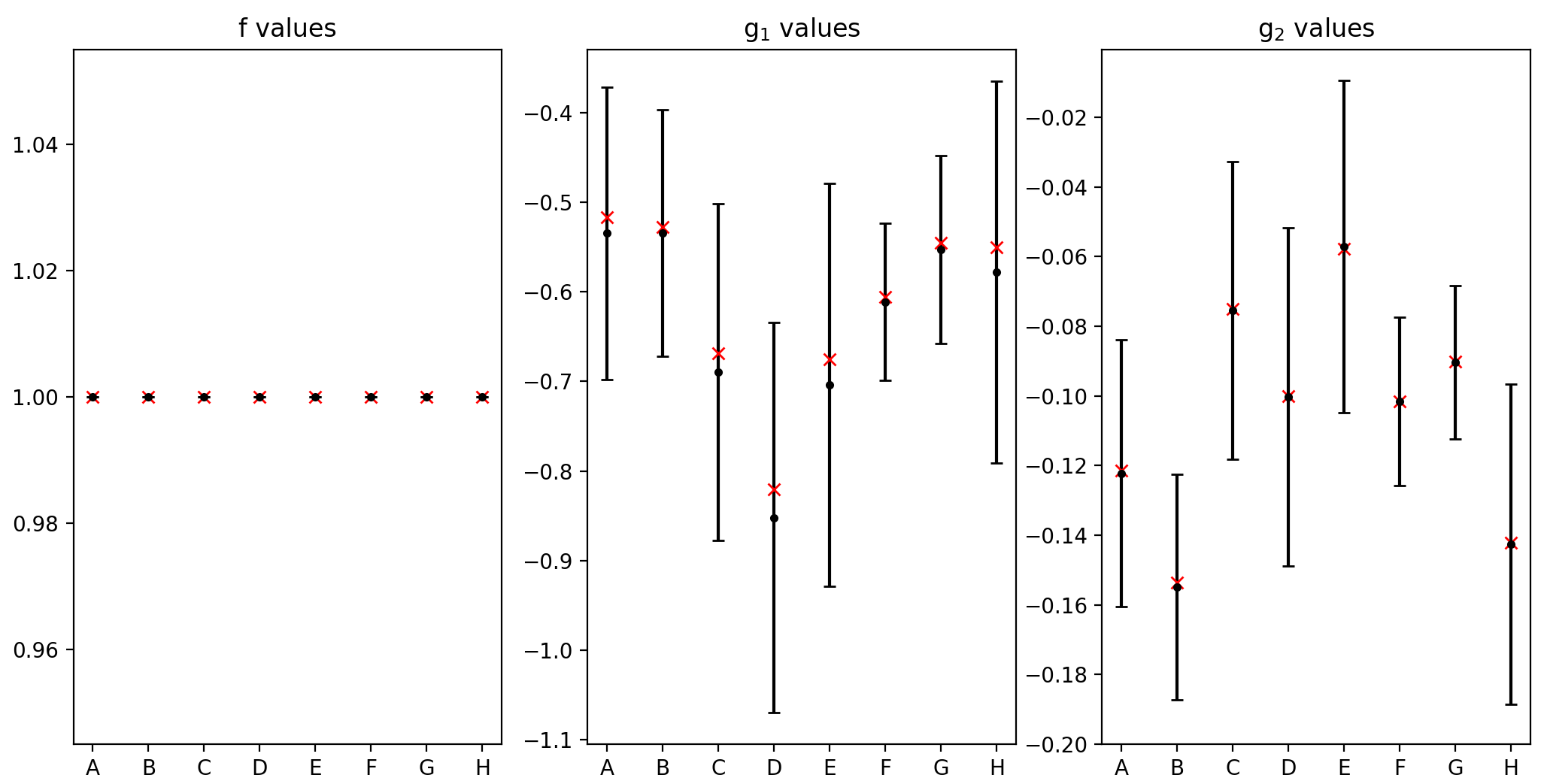}
	\vspace{-3ex}
	\caption{Local lens properties for image~1 in J2230 over all filter-band configurations in Table~\ref{tab:HO_labels_table}, analogous to Fig.~\ref{fig:cl0024_image_1}.}
	\label{fig:ptmatch_Image_1}
\end{figure}

\begin{figure}
	\centering
	\includegraphics[width=0.49\textwidth]{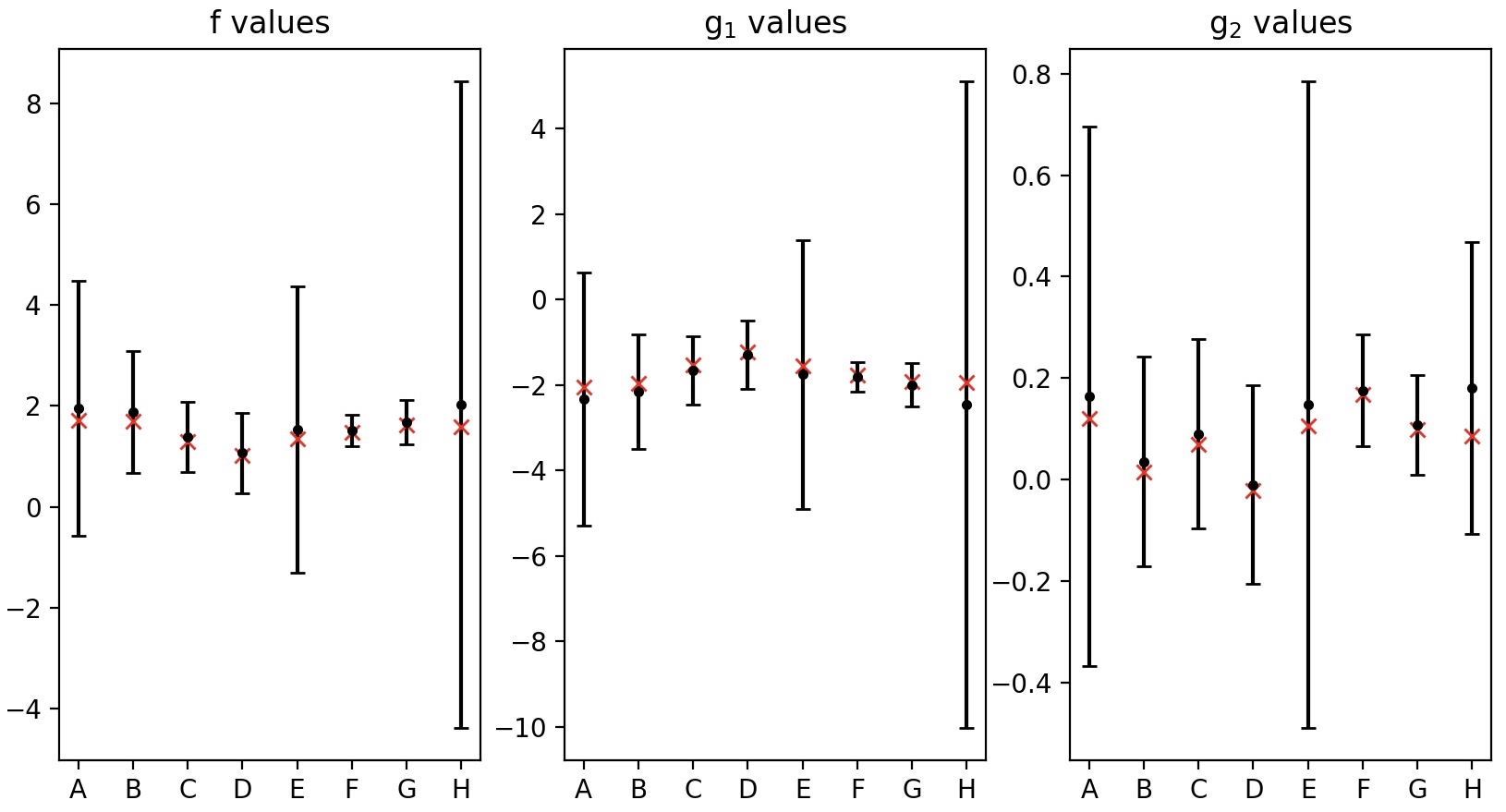}
	\vspace{-3ex}
	\caption{Same as Fig.~\ref{fig:ptmatch_Image_1} for image~2 in J2230.}
	\label{fig:ptmatch_Image_2}
\end{figure}

\begin{figure}
	\centering
	\includegraphics[width=0.49\textwidth]{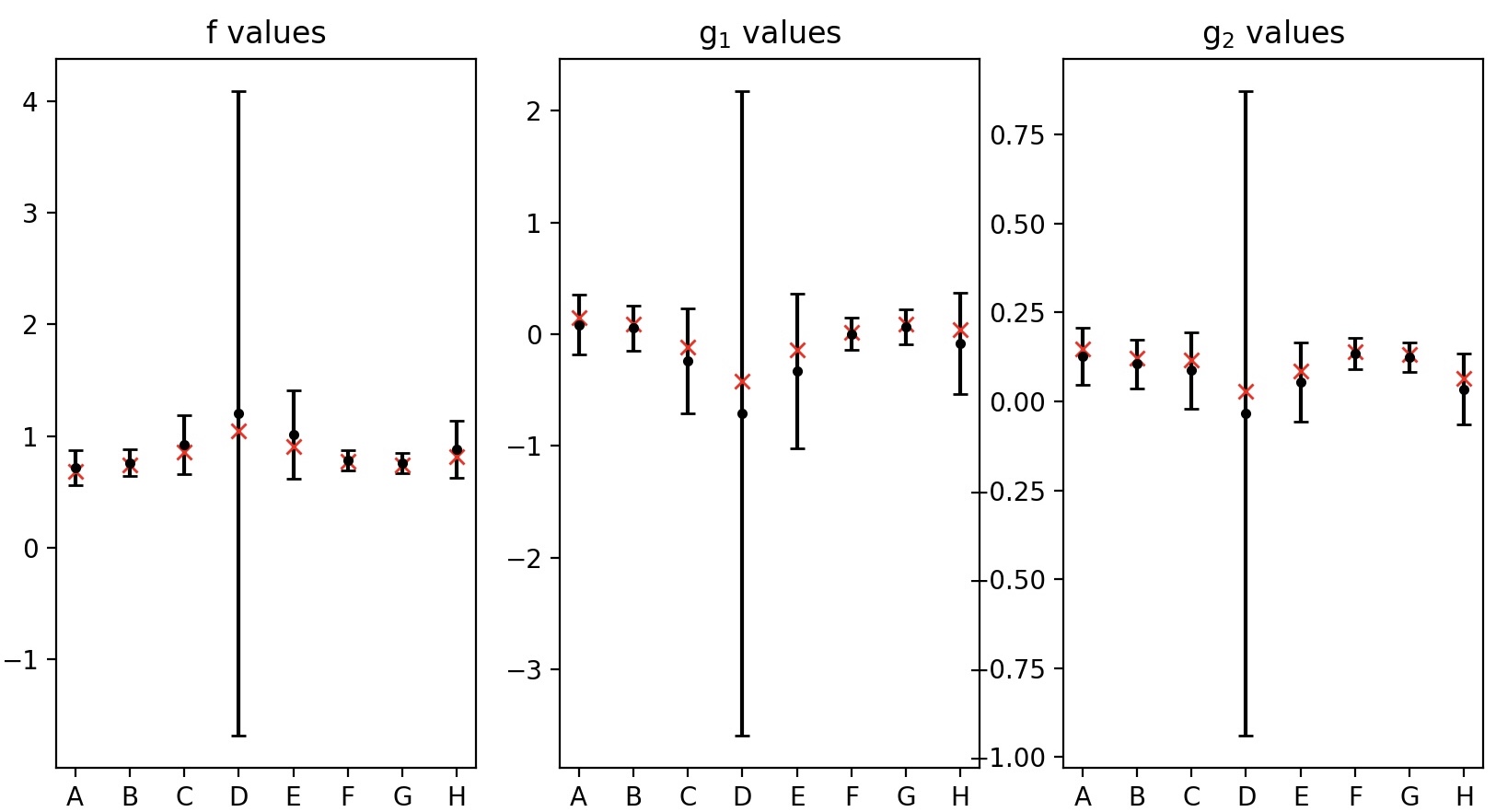}
	\vspace{-3ex}
	\caption{Same as Fig.~\ref{fig:ptmatch_Image_1} for image~3 in J2230.}
	\label{fig:ptmatch_Image_3}
\end{figure}

Using feature~4t, the local lens properties for all multiple images are shown in Figs.~\ref{fig:ptmatch_Image_1}, \ref{fig:ptmatch_Image_2}, and \ref{fig:ptmatch_Image_3}. 
(Due to different random realisations for the confidence bounds, the error bars compared to Fig.~\ref{fig:4t_4b} can differ.)
Table~\ref{tab:Hamilton_summary} summarises all values in detail.
As can be read off Table~\ref{tab:HO_labels_table}, all $\chi_{\rm red}^2$ values remain well under 1 and are in the regime of over-fitting.
For this case, over-fitting is not surprising, particularly because image~1 and image~2 are more or less reflected copies of each other, such that not all degrees of freedom offered by the linear transformation encoded by the local lens properties may be required for the mapping.  

Based on some very large error bars that are observed for the downsized versions of F606W and F814W and the colour image compared to the very robust results for the original data, it is possible that the additional processing of these observations introduces biases in the positions of the features. 
Therefore, any additional image pre-processing, for instance, denoising, should be handled with care in order not to bias the positions of the multiple images and its features. 
We also find that the confidence bounds for the local lens properties inferred in the F606W and F814W filter bands are all smaller, ranging from 12\% to 67\% of the ones obtained by manual feature extraction (at the worse resolution of the F110W filter band).
They are even smaller than the confidence bounds obtained using seven manually extracted features. 
Both results together corroborate that the resolution of the observation and its pre-processing has a non-negligible impact on the precision of the inferred local lens properties.

\begin{table}
		\caption{Flux ratios $F_2$ and $F_3$ between image~2 and 3 to image~1 in J2230 for the central bulge as given in Table~1 in \protect\cite{bib:Griffiths} for the filter bands used in our analyses.}
		\label{tab:HO_fluxes}		
		\begin{center}
			\begin{tabular}{c|c|c}
				\hline
				Waveband & $F_2$ & $F_3$ \\
				\hline
				F606W & 0.71 & 0.44 \\
				F814W & 0.79 & 0.36 \\
				F110W & 0.74 & 0.43 \\
				\hline                              
			\end{tabular}
		\end{center}
\end{table}

\begin{figure*}
	\centering
	\includegraphics[width=0.28\textwidth]{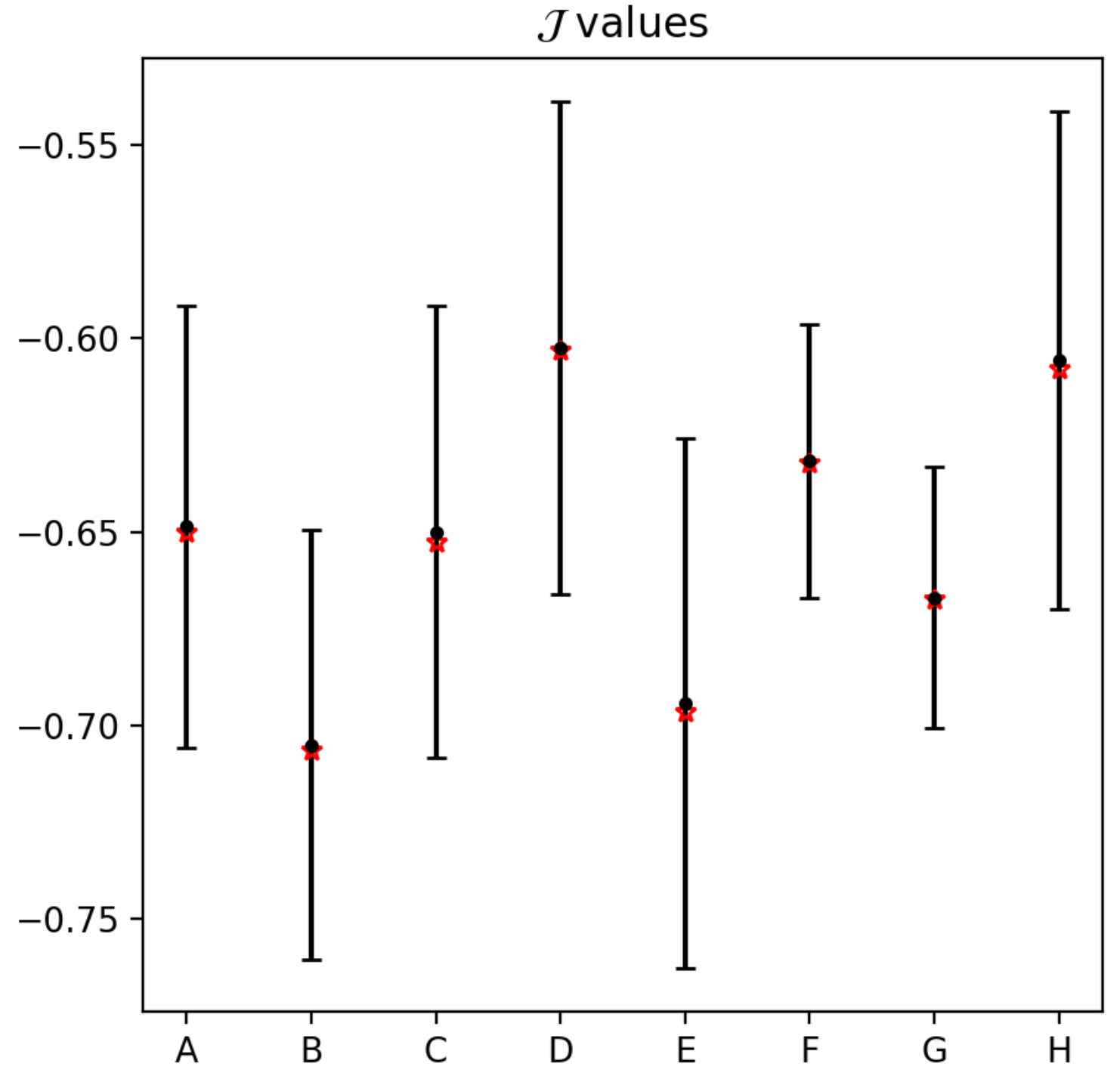}
	\hspace{4ex}
	\includegraphics[width=0.28\textwidth]{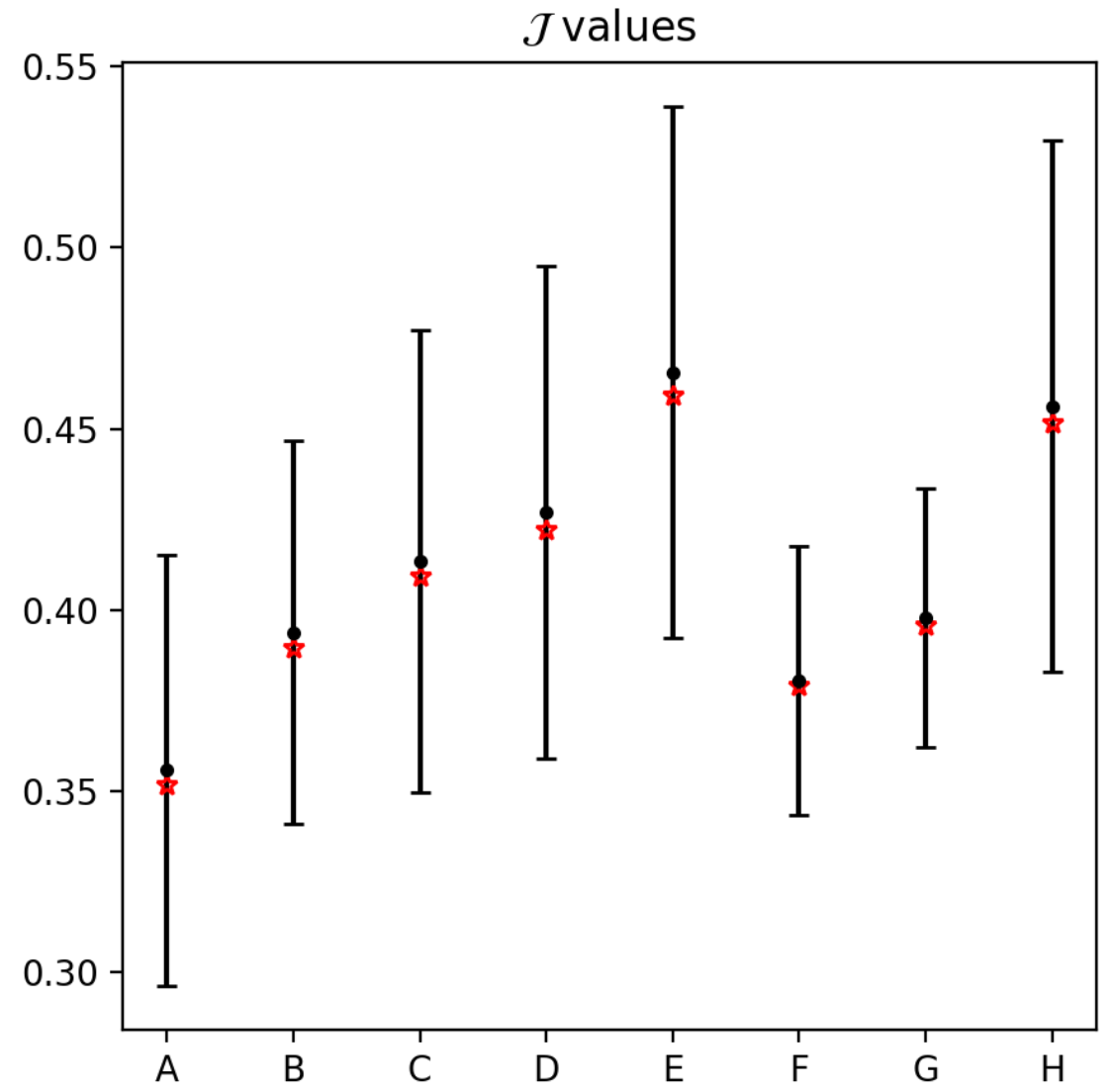}
	\caption{Magnification ratio $\mathcal{J}$, Eq.~\ref{eq:j}, for image~2 (left) and image~3 (right) with respect to image~1 in J2230 for all filter-band configurations of Table~\ref{tab:HO_labels_table}. Notation as in Fig.~\ref{fig:cl0024_image_1}.}
	\label{fig:HO_j}
\end{figure*}

At last, we plot the magnification ratios of images~2 and 3 to image~1, Eq.~\ref{eq:j}, inferred by \texttt{ptmatch} in Fig.~\ref{fig:HO_j}. 
Compared to the five-image configuration in CL0024, the confidence bounds on $\mathcal{J}_2$ and $\mathcal{J}_3$ are reduced for F606W and F814W, contributing to a stronger variation between the different configurations. 
As we see in Figs.~\ref{fig:4t_4b} and \ref{fig:ptmatch_Image_1}, the latter are mainly caused by the variations in image~1 over the different filter-band configurations.
All magnification ratios agree within their 68\% confidence bounds so that we find no hint for biases of the wavelength independence of strong gravitational lensing. 

From the measured magnitudes of the images contained in Table~1 of \cite{bib:Griffiths}, we obtain the flux rations for the three filter bands analysed in Table~\ref{tab:HO_fluxes}.
While there is some flux in feature~7 and also contributions from the other features, the central bulge was found to be the major source of emission in \cite{bib:Griffiths} and its observed flux was found in overall agreement with the magnification ratios as determined by \texttt{ptmatch}. 
The same is found for the magnification ratios obtained when using our automated feature extraction. 
Yet, all magnification ratios are systematically lower than the observed flux ratios except for $\mathcal{J}_3$ for F814W. 
This may be a hint of micro-lensing in image~1, supported by the detection of feature~4b. 
Assuming (locally) brighter than expected fluxes due to micro-lensing that are not incorporated in the \texttt{ptmatch} analysis, the flux ratios would be lower than the ones inferred from \texttt{ptmatch}. 
Alternatively, it is equally well possible that the parts outside the centre of the galaxy can explain the under-predicted magnification ratios. 
To follow up on that hypothesis, the same problem with the flux determination from a set of non-contiguous bright features as noted in Section~\ref{sec:cl0024} arises and is left for future work.

\begin{sidewaystable*} 
		\caption{ Synopsis of local lens properties, $\mathcal{J}_i, f_i,  g_{i,1}$, and $g_{i,2}$ of the five-image configuration in CL0024 for all filter-band configurations of Table~\ref{tab:CL0024_labels_table}. Results obtained by eye differ from those in \protect\cite{bib:Wagner_cluster} because \texttt{ptmatch} was re-run for our analyses, but results are based on the same input data and therefore statistically consistent.}
		\label{tab:CL0024_summary}
		\begin{center}
			\begin{tabular}{c|rr|rr|rr|rr|rr|rr|rr|rr}
				\hline
				& \multicolumn{2}{c|}{A} & \multicolumn{2}{c|}{B} &  \multicolumn{2}{c|}{C} &  \multicolumn{2}{c|}{D} & \multicolumn{2}{c|}{E} &  \multicolumn{2}{c|}{F} & \multicolumn{2}{c|}{G} & \multicolumn{2}{c}{H} \\
				& \multicolumn{2}{c|}{$\chi^{2}_{\rm red}$ = 1.35} & \multicolumn{2}{c|}{$\chi^{2}_{\rm red}$ = 1.73} & \multicolumn{2}{c|}{$\chi^{2}_{\rm red}$ = 0.752} & \multicolumn{2}{c|}{$\chi^{2}_{\rm red}$ = 0.15} & \multicolumn{2}{c|}{$\chi^{2}_{\rm red}$ = 0.62} & \multicolumn{2}{c|}{$\chi^{2}_{\rm red}$ = 0.69} & \multicolumn{2}{c|}{$\chi^{2}_{\rm red}$ = 1.41} & \multicolumn{2}{c}{$\chi^{2}_{\rm red}$ = 2.70} \\

				& Mean & Std & Mean & Std & Mean & Std & Mean & Std & Mean & Std & Mean & Std & Mean & Std & Mean & Std \\
				\hline
				$g_{1,1}$ & -0.06 & 0.02 & -0.06 & 0.02 & -0.04 & 0.03 & -0.06 & 0.03 & -0.03 & 0.02 & -0.06 & 0.03 & -0.07 & 0.3 & -0.06 & 0.02  \\
				$g_{1,2}$ & -0.65 & 0.05 & -0.63 & 0.05 & -0.63 & 0.05 & -0.58 & 0.06 & -0.60 & 0.05 & -0.59 & 0.06 & -0.57 & 0.07 & -0.60 & 0.05  \\
				\hline
				$\mathcal{J}_{2}$ & -0.42 & 0.05 & -0.43 & 0.04 & -0.41 & 0.05 & -0.44 & 0.05 & -0.43 & 0.05 & -0.40 & 0.04 & -0.39 & 0.04 & -0.41 & 0.05  \\
				$f_{2}$ & 1.76 & 0.41 & 1.88 & 0.39 & 1.82 & 0.44 & 2.50 & 1.25 & 2.22 & 0.69 & 2.00 & 0.67 & 2.23 & 2.56 & 1.75 & 0.41 \\
				$g_{2,1}$ & -0.32 & 0.11 & -0.28 & 0.1 & -0.31 & 0.12 & -0.37 & 0.19 & -0.41 & 0.16 & -0.38 & 0.14 & -0.34 & 0.33 & -0.44 & 0.12 \\
				$g_{2,2}$ & -2.27 & 0.50 & -2.44 & 0.51 & -2.40 & 0.55 & -3.24 & 1.68 & -2.87 & 0.87 & -2.73 & 0.89 & -3.06 & 4.08 & -2.37 & 0.50 \\
				\hline
				$\mathcal{J}_{3}$ & 0.72 & 0.06 & 0.70 & 0.05 & 0.73 & 0.06 & 0.72 & 0.06 & 0.70 & 0.06 & 0.70 & 0.06 & 0.72 & 0.06 & 0.68 & 0.06  \\
				$f_{3}$ & 0.99 & 0.10 & 0.94 & 0.08 & 0.96 & 0.10 & 0.90 & 0.09 & 0.91 & 0.09 & 0.89 & 0.09 & 0.89 & 0.09 & 0.90 & 0.08 \\
				$g_{3,1}$ & -0.48 & 0.05 & -0.50 & 0.04 & -0.49 & 0.05 & -0.50 & 0.04 & -0.48 & 0.04 & -0.50 & 0.04 & -0.50 & 0.04 & -0.49 & 0.04  \\
				$g_{3,2}$ & 0.02 & 0.09 & 0.05 & 0.09 & 0.03 & 0.09 & 0.13 & 0.09 & 0.10 & 0.08 & 0.10 & 0.09 & 0.15 & 0.09 & 0.06 & 0.09   \\
				\hline
				$\mathcal{J}_{4}$ & -0.73 & 0.07 & -0.73 & 0.06 & -0.70 & 0.07 & -0.77 & 0.08 & -0.74 & 0.07 & -0.69 & 0.07 & -0.71 & 0.07 & -0.67 & 0.07 \\
				$f_{4}$ &  1.50 & 0.27 & 1.60 & 0.24 & 1.45 & 0.26 & 1.86 & 0.42 & 1.67 & 0.31 & 1.55 & 0.30 & 1.71 & 0.42 & 1.36 & 0.24 \\
				$g_{4,1}$ & 0.21 & 0.07 & 0.20 & 0.07 & 0.25 & 0.07 & 0.36 & 0.12 & 0.28 & 0.08 & 0.28 & 0.09 & 0.37 & 0.13 & 0.22 & 0.07   \\
				$g_{4,2}$ & -1.65 & 0.21 & -1.76 & 0.21 & -1.65 & 0.20 & -1.97 & 0.36 & -1.83 & 0.26 & -1.79 & 0.25 & -1.92 & 0.38 & -1.65 & 0.19 \\
				\hline
				$\mathcal{J}_{5}$ & 0.19 & 0.03 & 0.18 & 0.03 & 0.18 & 0.03 & 0.19 & 0.03 & 0.18 & 0.03 & 0.15 & 0.03 & 0.17 & 0.03 & 0.18 & 0.03 \\
				$f_{5}$ & -0.53 & 0.06 & -0.48 & 0.05 & -0.51 & 0.05 & -0.50 & 0.06 & -0.47 & 0.06 & -0.43 & 0.06 & -0.46 & 0.06 & -0.48 & 0.06 \\
				$g_{5,1}$ & 0.07 & 0.07 & 0.07 & 0.06 & 0.03 & 0.07 & 0.08 & 0.08 & 0.11 & 0.07 & 0.09 & 0.08 & 0.13 & 0.08 & 0.07 & 0.07  \\
				$g_{5,2}$ & -0.42 & 0.10 & -0.43 & 0.08 & -0.38 & 0.10 & -0.33 & 0.11 & -0.40 & 0.10 & -0.39 & 0.11 & -0.37 & 0.11 & -0.43 & 0.10 \\
				\hline                                                          
			\end{tabular}
		\end{center}
\end{sidewaystable*}
	
\begin{sidewaystable*} 
		\caption{ Synopsis of local lens properties, $\mathcal{J}_i, f_i,  g_{i,1}$, and $g_{i,2}$, of the three-image configuration in J2230 for all filter-band configurations of Table~\ref{tab:HO_labels_table}. Results obtained by eye differ from those in \protect\cite{bib:Griffiths} because \texttt{ptmatch} was re-run for our analyses and the reference image was exchanged. Results are based on the same input data and therefore statistically consistent. The order is rearranged to have the filter bands and their down-sized versions next to each other for easier comparison.}
		\label{tab:Hamilton_summary}		
		\begin{center}
			\begin{tabular}{c|rr|rr|rr|rr|rr|rr|rr|rr}
				\hline
				& \multicolumn{2}{c|}{A} &  \multicolumn{2}{c|}{B} & \multicolumn{2}{c|}{F} & \multicolumn{2}{c|}{D} & \multicolumn{2}{c|}{G} & \multicolumn{2}{c|}{E} & \multicolumn{2}{c|}{C} & \multicolumn{2}{c}{H} \\
				& \multicolumn{2}{c|}{$\chi^{2}_{\rm red}$ = 0.06} & \multicolumn{2}{c|}{$\chi^{2}_{\rm red}$ = 1.11} & \multicolumn{2}{c|}{$\chi^{2}_{\rm red}$ = 0.20} & \multicolumn{2}{c|}{$\chi^{2}_{\rm red}$ = 0.12} & \multicolumn{2}{c|}{$\chi^{2}_{\rm red}$ = 0.42} & \multicolumn{2}{c|}{$\chi^{2}_{\rm red}$ = 0.73} & \multicolumn{2}{c|}{$\chi^{2}_{\rm red}$ = 0.87} & \multicolumn{2}{c}{$\chi^{2}_{\rm red}$ = 0.22} \\
				& Mean & Std & Mean & Std & Mean & Std & Mean & Std & Mean & Std & Mean & Std & Mean & Std & Mean & Std\\
				\hline
				$g_{1,1}$ & -0.53 & 0.16 & -0.53 & 0.14 & -0.61 & 0.09 & -0.85 & 0.22 & -0.55 & 0.10 & -0.70 & 0.23 & -0.70 & 0.19 & -0.58 & 0.21 \\
				$g_{1,2}$ & -0.12 & 0.04 & -0.15 & 0.03 & -0.10 & 0.02 & -0.10 & 0.05 & -0.09 & 0.02 & -0.06 & 0.05 & -0.08 & 0.04 & -0.14 & 0.05  \\
				\hline
				$\mathcal{J}_{2}$ & -0.65 & 0.06 & -0.71 & 0.06 & -0.63 & 0.04 & -0.60 & 0.06 & -0.67 & 0.03 & -0.69 & 0.07 & -0.65 & 0.06 & -0.61 & 0.06 \\
				$f_{2}$ & 1.95 & 2.53 & 1.87 & 1.21 & 1.52 & 0.31 & 1.07 & 0.80 & 1.68 & 0.44 & 1.54 & 2.84 & 1.39 & 0.92 & 2.03 & 6.41 \\
				$g_{2,1}$ & -2.34 & 2.96 & -2.16 & 1.35 & -1.81 & 0.34 & -1.30 & 0.80 & -2.00 & 0.50 & -1.75 & 3.14 & -1.66 & 1.05 & -2.46 & 7.57 \\
				$g_{2,2}$ & 0.16 & 0.53 & 0.04 & 0.21 & 0.18 & 0.11 & -0.01 & 0.20 & 0.11 & 0.10 & 0.15 & 0.64 & 0.09 & 0.23 & 0.18 & 0.29 \\
				\hline
				$\mathcal{J}_{3}$ & 0.36 & 0.06 & 0.39 & 0.05 & 0.38 & 0.04 & 0.43 & 0.07 & 0.40 & 0.04 & 0.47 & 0.07 & 0.41 & 0.06 & 0.46 & 0.07 \\
				$f_{3}$ & 0.72 & 0.16 & 0.76 & 0.12 & 0.78 & 0.09 & 1.20 & 2.88 & 0.76 & 0.09 & 1.02 & 0.39 & 0.93 & 0.28 & 0.88 & 0.25 \\
				$g_{3,1}$ & 0.08 & 0.27 & 0.06 & 0.20 & 0.00 & 0.15 & -0.71 & 2.88 & 0.07 & 0.16 & -0.33 & 0.69 & -0.26 & 0.50 & -0.08 & 0.45 \\
				$g_{3,2}$ & 0.13 & 0.08 & 0.10 & 0.07 & 0.13 & 0.04 & -0.03 & 0.91 & 0.12 & 0.04 & 0.05 & 0.11 & 0.08 & 0.11 & 0.03 & 0.10 \\
				\hline                                                          
			\end{tabular}
		\end{center}
\end{sidewaystable*}

\section{Discussion and summary of results}
\label{sec:discussion}

In summary, this part of the paper series focuses on the identification of robust features within multiple images instead of the local lens reconstruction method as the previous parts.
As such, it yields valuable information for other lens reconstruction approaches as well. 

We showed that standard techniques exploiting the physical principles of the data acquisition, for instance implemented in \texttt{SExtractor}, are still adequate to set up an efficient feature extraction.
A calibration on one representative example case scales well across different filter bands, resolutions, and signal-to-noise environments.
Enhanced deblending and noise reduction, like \cite{bib:Melchior3}, may increase the number of detected features in crowded and noisy environments.
However, it comes at a higher computational and run-time cost and, using surface-density brightness profiles and noise models to deblend and denoise objects, it may also bias the feature extraction.
Thus, it should only be used if the simple attempt does not yield the required outcome. 
For the two multiple image configurations analysed here, the cost-saving method found less features than manual identification, but detected the salient ones that led to comparable or even smaller sizes of confidence bounds for the inferred local lens properties.

Concerning the feature positions, we can combine the findings of \cite{bib:Wagner7} that higher-order lensing effects like flexion can mostly be neglected for galaxy-scale multiple images in galaxy-cluster scale lenses with the findings of Section~\ref{sec:calibration} that robust features have peak positions within 1--2 pixel distance to their centre of light. 
Consequently, using the peak position as the theoretically motivated coordinates (see Section~\ref{sec:robust_features}) or using the centre of light as the coordinates robust against noise should not lead to strongly deviating results in the lens reconstruction. 
As detailed in Section~\ref{sec:calibration}, the choice which one is better suited depends on the signal-to-noise ratio of the observation under analysis and the desired lens reconstruction precision. 

Depending on the pre-processing of the data, biases in the positions can be introduced which do not increase the $\chi^2_{\rm red}$.
This implies that the latter should not be taken as the sole goodness-of-fit measure to evaluate whether the lens reconstruction is physically meaningful.
As an example detailed in Section~\ref{sec:hamiltons_object}, the local lens properties obtained from the images in the downsized versions of F606W and F814W have a $\chi^2_{\rm red} < 1$, but larger confidence bounds than other local lens properties.

Setting up persistence diagrams to track detected objects over increasing significance levels increases the run-time and computational costs compared to a single thresholding. 
At the same time, extracting only features fulfiling the persistence criteria outlined in Section~\ref{sec:robust_feature_extraction}, noisy peak positions and a manually undetected feature could both be successfully identified, alleviating or at least detecting potential biases caused by the pre-processing.
Specifically for our approach to reconstruct local lens properties using \texttt{ptmatch} (or, the analytic approximation to the critical curve detailed in \cite{bib:Wagner1}), the automated feature extraction allows for a standardised evaluation by defining quantitative selection criteria for features to be matched. It even succeeded in finding a sufficient amount of features when a manual approach was not successful.

While inconsistencies can show up for an incorrect matching between features across multiple images clearly pointing at the matching error, for instance shown in \cite{bib:Wagner_quasar}, ambiguities to match available features can also lead to viable results for all matching possibilities. 
Section~\ref{sec:hamiltons_object} details one such case for the two possible feature~4 in image~1 of Hamilton's Object.
Further examples for ambiguous feature matchings are Abell~3827, see, for instance, \cite{bib:Chen}, in which the present matching is incompatible to standard leading-order lensing theory similar to the case in \cite{bib:Wagner_quasar} or the matching of features in the giant arc in WHL0137-08 according to \cite{bib:Welch}.
They only show one possible matching and this only maps parts of the giant arc onto each other. 
Since the resulting lens reconstructions can differ for different matchings, conclusions about properties of (dark) matter should be treated with caution in such cases. 
For \texttt{ptmatch} we showed that translations of feature coordinates are only related to the corresponding local lens property constrained by these coordinates. 
Hence, coordinate changes and entailed changes in local lens properties are directly linked and the impact of the former on the latter is intuitively understandable.

\section{Conclusion}
\label{sec:conclusion}

This part of the paper series focused on the development of an automated feature extraction approach to robustly and quantitatively identify bright, elliptical intensity features like star-forming regions in multiple images of strong gravitationally lensed background galaxies. 
Our approach is based on publicly available software, \texttt{sep}, adapted and calibrated to our task as detailed in Section~\ref{sec:robust_feature_extraction}. 
The technical benefits for our lens reconstruction approach and others are summarised in Section~\ref{sec:discussion}, which also gives recommendations for the detection and the usage of brightness features to infer (local) lens properties based on the results found for two representative multiple image configurations detailed in Section~\ref{sec:applications}. 

Most importantly, we found that, for brightness features within multiple images or for featureless multiple images generated by a galaxy-cluster scale strong gravitational lens, the fact that higher-order lensing effects like flexion are negligible in most cases (\cite{bib:Wagner7}), puts the maximum intensity position and the centre of light only 1--2 pixel apart from each other. 
As a consequence, basing lens reconstructions on either of them has a small impact on the lens reconstruction. 
For our lens reconstruction approach, using either of them leads to indistinguishable local lens properties within their confidence bounds. 
Furthermore, we showed that any change in the feature positions directly and understandably shifts the corresponding local lens property of the respective multiple image. 
The impact for other lens reconstruction approaches may not be so simple to single out. 

How well the feature extraction performs in further cases remains to be investigated. 
Yet, as we saw in our manual cross-checks for the two examples considered in this paper, the persistence diagrams that our feature extraction is based on yield the necessary information on occurring anomalies to exclude these cases. 
In 48 robust feature detections that our examples required, only one needed a manual follow-up due to the importance of that feature which was located in a low signal-to-noise regime. 
With the results we found, even that case can be automatically treated in the future. 

The much more problematic issue is the automation of the feature matching which currently seems impossible in principle to avoid making any a priori assumption about the lens when matching the features.
In that sense, every lens may be unhappy in its own way and require manual adjustment in their feature matching but feature extraction, as shown here, is easily automated. 

As the automated feature extraction can improve the precision of the inferred local lens properties, maximally over 30\% decrease in the confidence bounds for the cases considered here, it may become possible in the future to detect deviations from the wavelength independence of strong gravitational lensing. 
As estimated in \cite{bib:Er}, plasma lensing effects may only shift image positions of the order of tens of milli-arcseconds, so that its impact is negligible at the current resolution of telescopes and particularly at the observation wavelengths considered here. 
A more likely impact is caused by the dust attenuation affecting the overall observed flux of a multiple image in the blue optical wavelengths and micro-lensing due to granular matter densities close to the critical curve as simulated, for instance, in \cite{bib:Diego}.

By construction of our approach, the inferred magnification ratios have the smallest confidence bounds and can thus be compared to the measured flux ratios between pairs of multiple images over multiple filter bands. 
These comparisons allow us to investigate the influence of dust or (transient) micro-lensing on top of the galaxy-cluster scale strong gravitational lensing generating the multiple images under analysis. 
For the five-image configuration in CL0024 and the triple-image configuration in J2230, all local lens properties were in agreement with each other for all filter bands analysed.
A rough comparison to the measured flux ratios for the triple-image configuration in J2230 also showed an overall agreement. 
While model-based lens reconstructions require a global lens mass reconstruction for the entire area on the sky where the multiple images are located to determine the magnification ratios of the multiple images of interest, our approach constrains them without making any additional assumption about the global lens-mass density profile and therefore requires much less run-time and resources to arrive at the same result. 
Tests for anomalous flux ratios of multiple images are available in about a second and can be used to inform lens model reconstructions about additional attenuation or micro-lenses or be used as a standalone to scan a lot of multiple image configurations for flux ratio anomalies that may not be explainable in terms of these two known contaminations. 
For anomalous flux ratios in individual filter bands, deviations from cold dark matter or non-linear lensing effects accumulating deflections along the line of sight can be considered, while deviations between filter bands are expected to be caused by baryonic effects.

\section*{Acknowledgements}

We would like to thank Nicolas Tessore for his image mapping software, \texttt{pmatch}, and Kyle Barbary for his Python code of SExtractor, \texttt{sep}, which both were crucial to this project.
This paper uses data from the Hubble Legacy Archive, which is a collaboration between the Space Telescope Science Institute (STScI/NASA), the Space Telescope European Coordinating Facility (ST-ECF/ESAC/ESA) and the Canadian Astronomy Data Centre (CADC/NRC/CSA).

\section{Data availability}

All HST observations used in this work are publicly available in the Hubble Legacy Archive (HL). 
\texttt{sep} is an open source software wich can be downloaded here:
\url{https://sep.readthedocs.io/en/v1.1.x/}.
The same applies for \texttt{ptmatch}, \url{https://github.com/ntessore/imagemap}, and our own code, which is available here:
\url{https://github.com/joycelin1123/SEP-Automated-Feature-Detection}. 
 


\bibliographystyle{mnras}
\bibliography{references} 



\appendix

\section{Magnification ratios for CL0024}
\label{app:cl0024_magnification_ratios}
Fig.~\ref{fig:CL0024_j} shows the magnification ratios between images~2--5 and image~1 as determined by \texttt{ptmatch} as further detailed in Section~\ref{sec:cl0024}.

\begin{center}
	\includegraphics[width=0.28\textwidth]{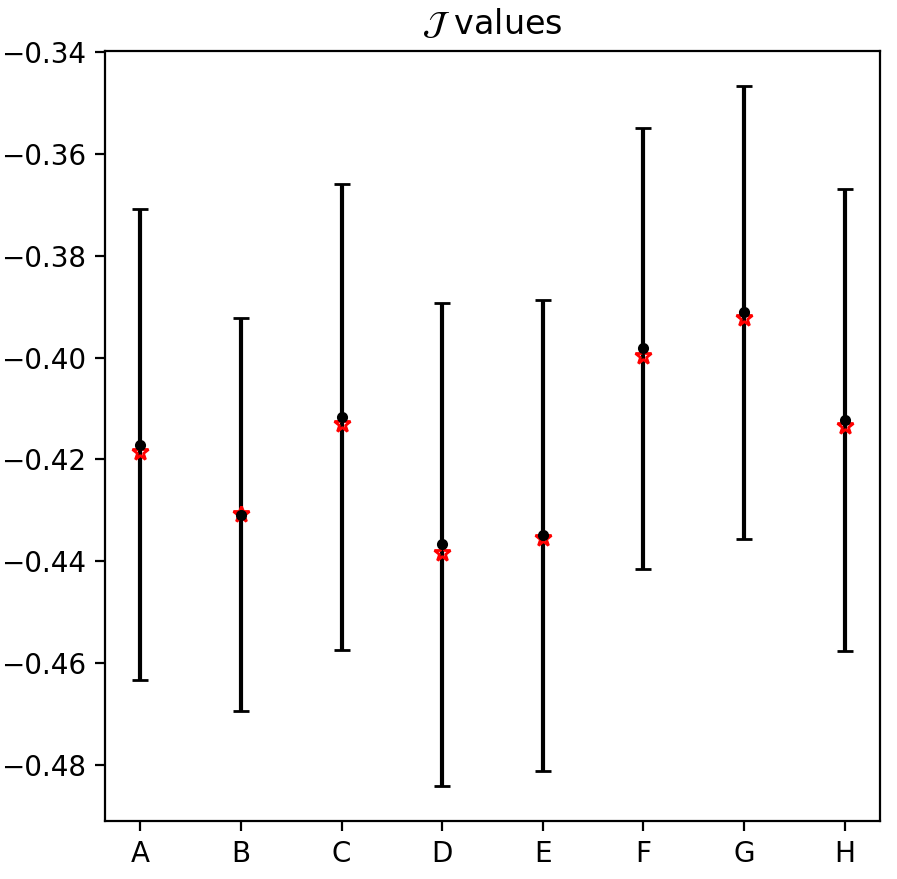} \hspace{4ex}
	\includegraphics[width=0.28\textwidth]{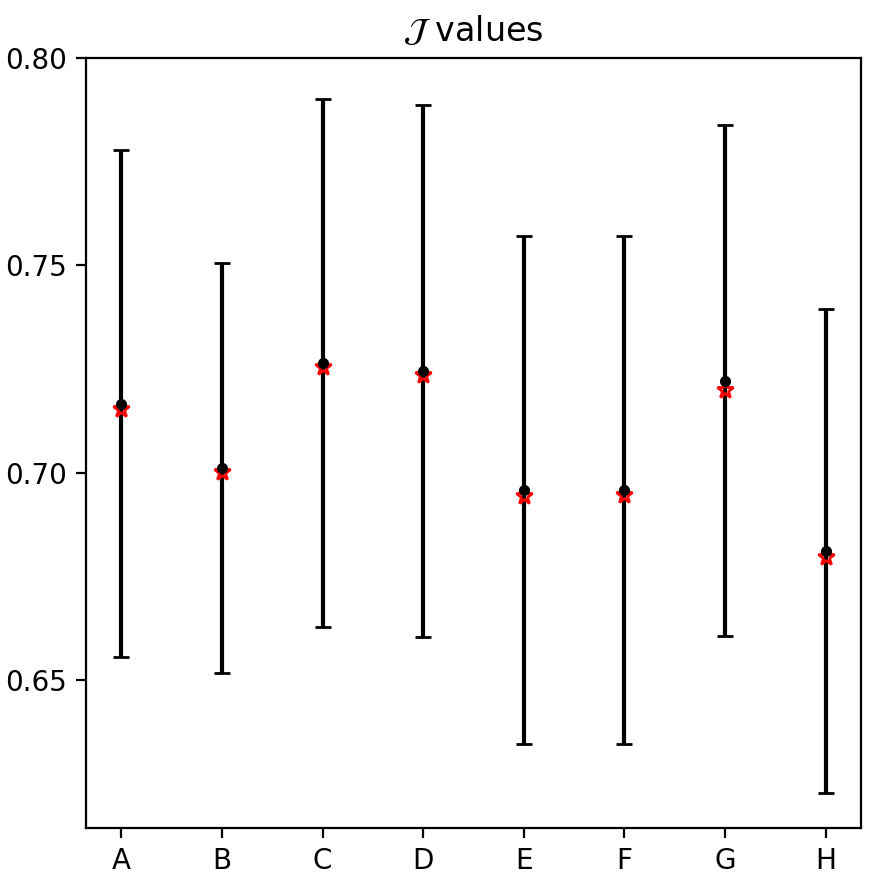}
\\[1ex]
	\includegraphics[width=0.28\textwidth]{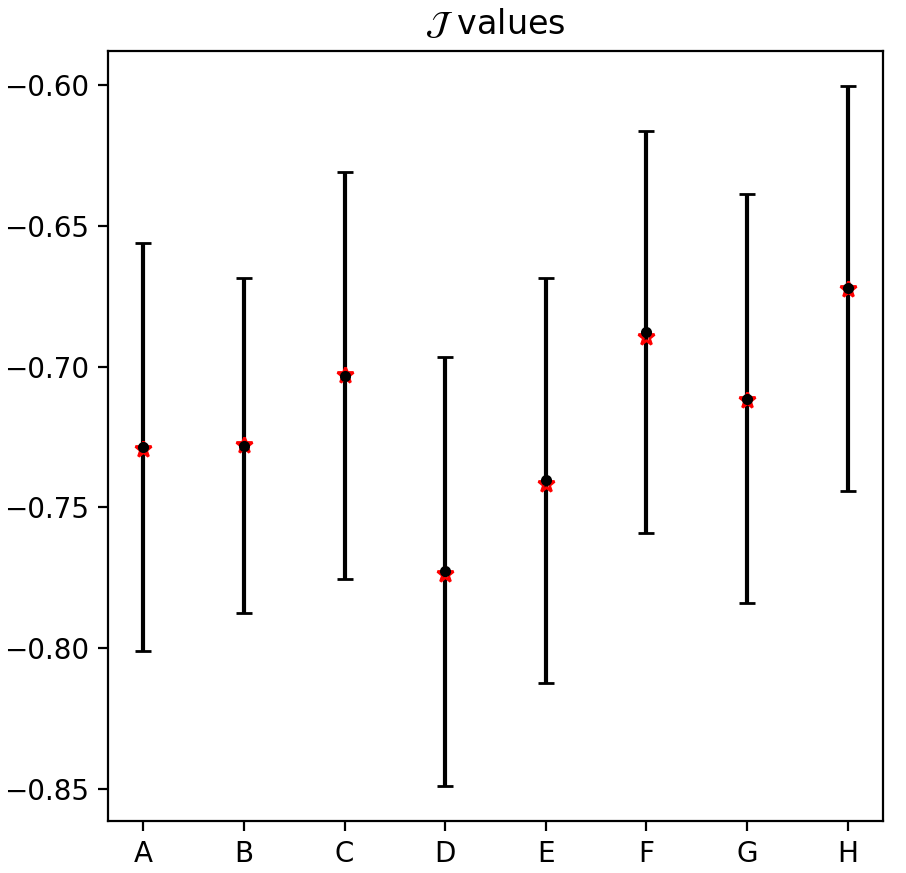} \hspace{4ex}
	\includegraphics[width=0.28\textwidth]{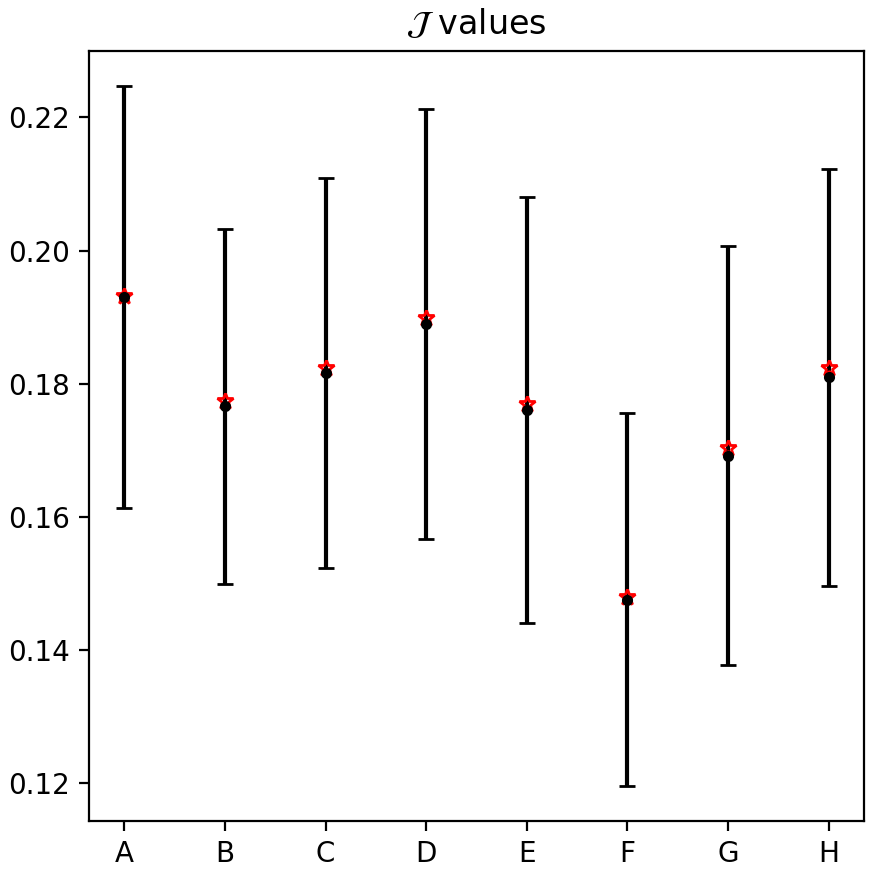}
	\\
\textbf{Fig.~A1.} Magnification ratio $\mathcal{J}$, Eq.~\ref{eq:j}, from top to bottom: for image~2, image~3, image~4, and image~5 with respect to image~1 in CL0024 for all filter bands of Table~\ref{tab:CL0024_labels_table} in the same notation as in Fig.~\ref{fig:cl0024_image_1}.
	\label{fig:CL0024_j}
\end{center}


\section{Dependency of local lens properties on feature selection for image~2 and 3 in J2230}
\label{app:4t_4b}

Fig.~\ref{fig:ptmatch_Image_2_3} shows the local lens properties for images~2 and 3 in J2230 for both features~4t and 4b as further detailed in Section~\ref{sec:hamiltons_object}.

\begin{center}
	\centering
	\includegraphics[width=0.49\textwidth]{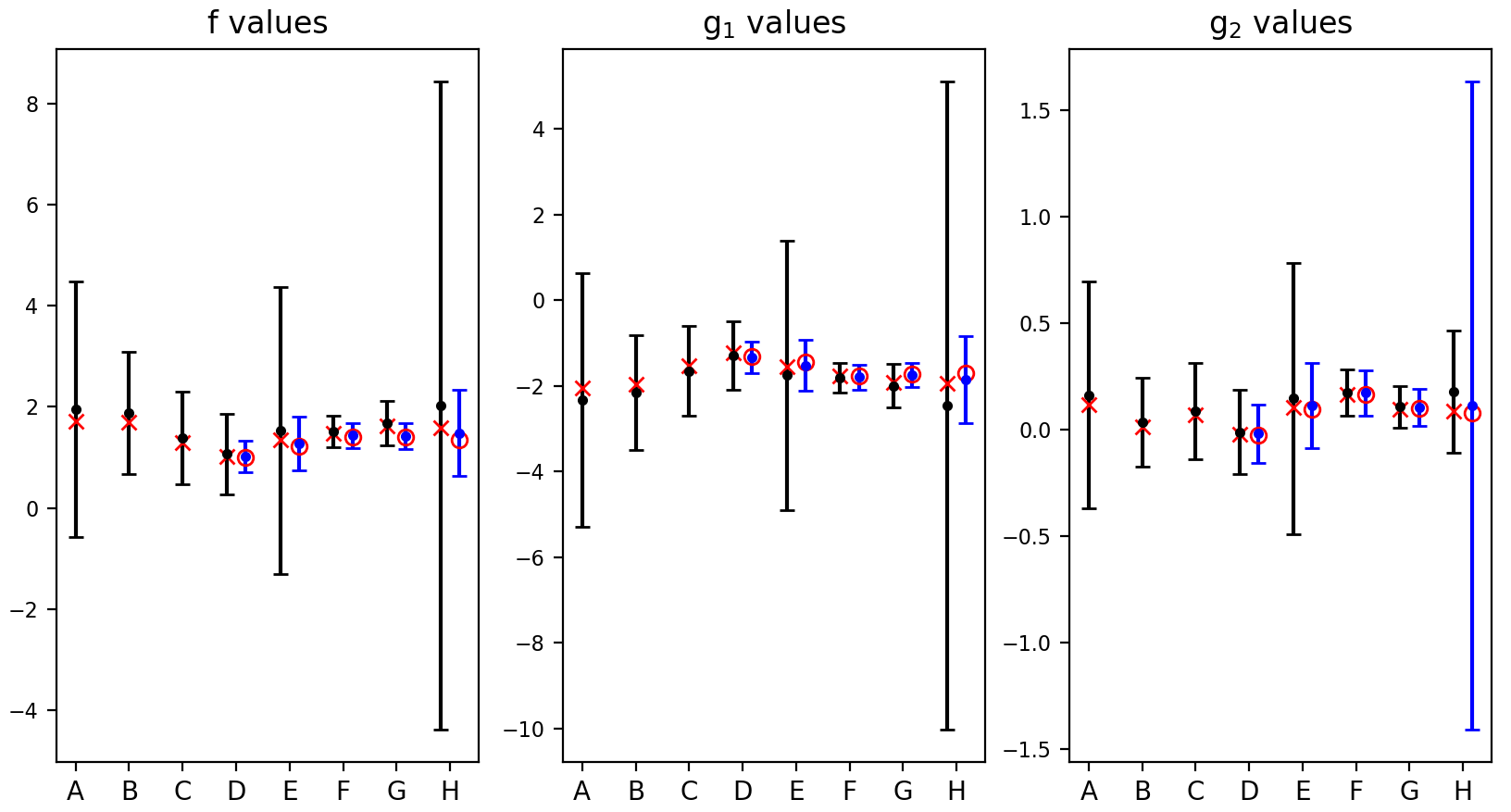}
	\includegraphics[width=0.49\textwidth]{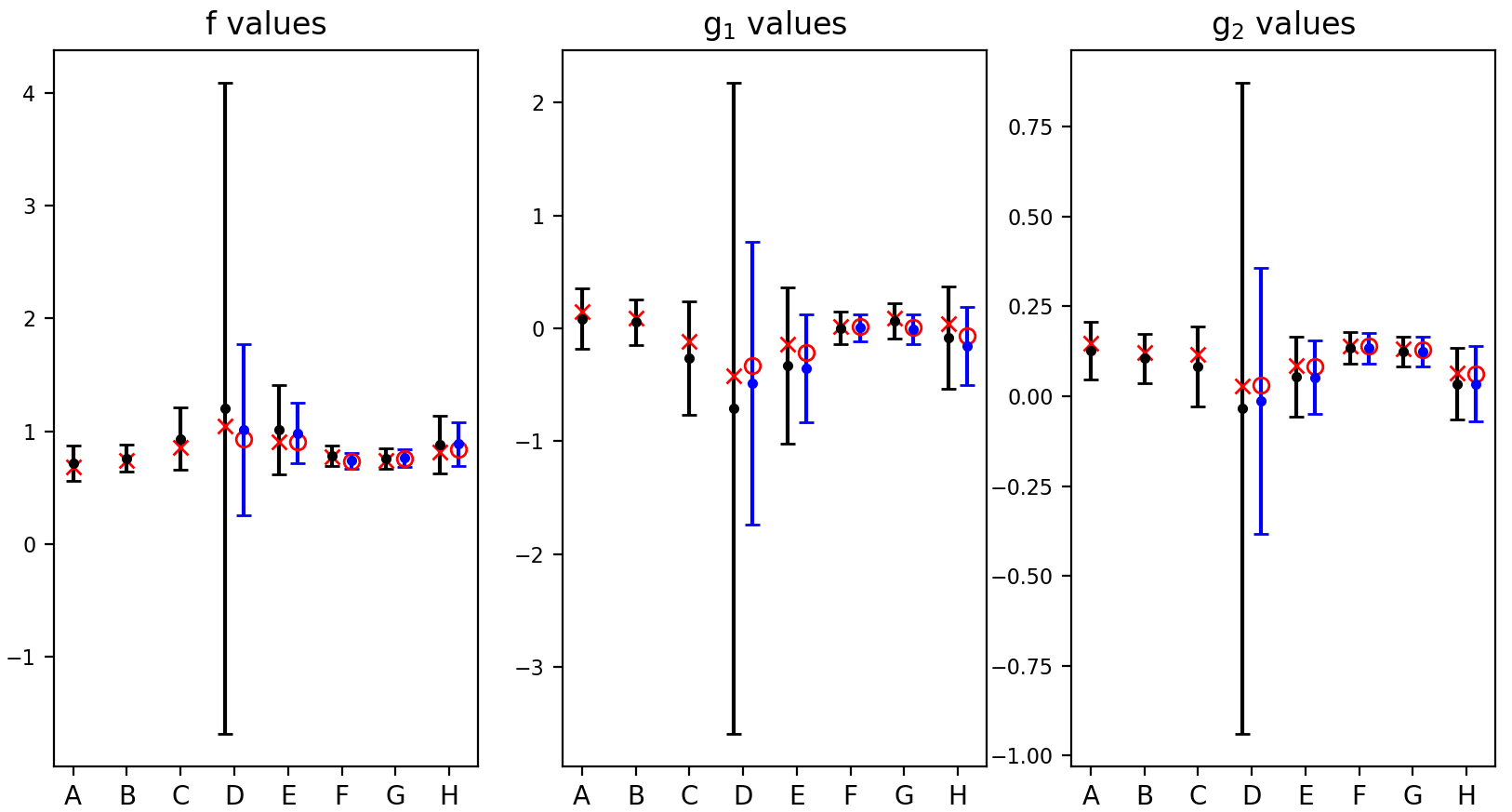}
	\textbf{Fig.~B1.} Local lens properties for image~2 (top) and image~3 (bottom) in J2230 for all filter-map configurations summarised in Table~\ref{tab:HO_labels_table} using feature~4t (black lines) and 4b (blue lines) as feature~4 in image~1.  
	\label{fig:ptmatch_Image_2_3}
\end{center}

\bsp	
\label{lastpage}
\end{document}